\documentclass[journal]{IEEEtran}
\ifCLASSINFOpdf

\else

\fi

\usepackage{graphicx}
\usepackage{hyperref}
\usepackage{indentfirst}
\usepackage{amsmath,amssymb,amsfonts}
\usepackage{algorithm}
\usepackage{multirow}
\usepackage{booktabs}
\usepackage{makecell}
\usepackage{url}
\usepackage{cite}
\usepackage{graphicx}
\usepackage{amsmath}
\usepackage{color}
\usepackage{amssymb}
\usepackage{subfigure}
\usepackage{multirow}
\usepackage{fancyhdr}
\usepackage{epic}
\usepackage{eepic}
\usepackage{amsmath}
\usepackage{threeparttable}
\usepackage{amscd}
\usepackage{here}
\usepackage{graphicx}
\usepackage{lscape}
\usepackage{tabularx}
\usepackage{longtable}
\usepackage{comment}

\usepackage{array}

\usepackage{soul, color}

\soulregister{\cite}7 % 注册\cite命令
\soulregister{\citep}7 % 注册\citep命令
\soulregister{\citet}7 % 注册\citet命令
\soulregister{\ref}7 % 注册\ref命令
\soulregister{\pageref}7 % 注册\pageref命令

\usepackage{lipsum} % To generate some text for the article via \lipsum

\usepackage{tikz,xcolor,hyperref}

\definecolor{lime}{HTML}{A6CE39}
\DeclareRobustCommand{\orcidicon}{%
	\begin{tikzpicture}
		\draw[lime, fill=lime] (0,0)
		circle [radius=0.16]
		node[white] {{\fontfamily{qag}\selectfont \tiny ID}};
		\draw[white, fill=white] (-0.0625,0.095)
		circle [radius=0.007];
	\end{tikzpicture}
	\hspace{-2mm}
}

\foreach \x in {A, ..., Z}{%
	\expandafter\xdef\csname orcid\x\endcsname{\noexpand\href{https: //orcid.org/\csname orcidauthor\x\endcsname}{\noexpand\orcidicon}}
}

\usepackage{multirow} %关于表格
\usepackage{booktabs}
\usepackage{makecell}

\usepackage{algorithm}
\usepackage{algpseudocode}
\usepackage{amsmath}
\usepackage{color}
\usepackage{amssymb}

\begin{document}

\title{A Blockchain-based Secure Storage Scheme for Medical Information}

\author{
Zhijie Sun \orcidA{}
Dezhi Han \orcidB{}, %\textit{Member, IEEE},
Dun Li \orcidC{}, %\textit{Student Member, IEEE},
Xiangsheng Wang \orcidD{},
Chin-Chen Chang,
Zhongdai Wu

\thanks{
Zhijie Sun, Dezhi Han, and Xiangsheng Wang are with the College of Information Engineering at Shanghai Maritime University, China. E-mail: S19117157825@163.com.
}

\thanks{
Dun Li is with the College of Information Engineering at Shanghai Maritime University, China and Telecom SudParis, IMT, Institut Polytechnique de Paris, France.
E-mail: lidunshmtu@outlook.com.
(Corresponding Author)
}

\thanks{
Chin-Chen Chang is with the Department of Information Engineering and Computer Science, Feng Chia University, Taichung, Taiwan.
E-mail: alan3c@gmail.com.
} 	

\thanks{
Zhongdai Wu is with theCOSCO Shipping Technology Co., Shanghai, China.
E-mail: wzd@cnshipping.com.
} 	

}
\maketitle

\begin{abstract}
%医疗数据涉及大量个人信息，是高度隐私敏感的。
%大数据时代使得医疗信息化程度不断加深，因此，安全而准确的存储医疗信息至关重要。
%然而，当前的医疗信息存在隐私泄露的风险和共享困难。
%为了解决以上问题，本文提出了一种基于Hyperledger fabric和基于属性的访问控制(ABAC)框架的医疗信息安全存储方案。
%该方案首先利用基于属性的访问控制，可以对医疗信息进行动态和细粒度的访问，然后将医疗信息存入到区块链中，通过制定相应的智能合约，可以保证医疗信息的安全和不可篡改。
%此外，本方案还结合IPFS技术来缓解区块链的存储压力。
%实验表明，本文所提出的将属性的访问控制和区块链技术相结合的方案不仅能够保证医疗信息的安全存储和完整性，而且在存取医疗信息时具有较高的吞吐量。
Medical data involves a large amount of personal information and is highly privacy sensitive.
In the age of big data, the increasing informatization of healthcare makes it vital that medical information is stored securely and accurately.
However, current medical information is subject to the risk of privacy leakage and difficult to share.
To address these issues, this paper proposes a healthcare information security storage solution based on Hyperledger Fabric and the Attribute-Based Access Control (ABAC) framework.
The scheme first utilizes attribute-based access control, which allows dynamic and fine-grained access to medical information, and then stores the medical information in the blockchain, which can be secured and tamper-proof by formulating corresponding smart contracts.
In addition, this solution also incorporates IPFS technology to relieve the storage pressure of the blockchain.
Experiments show that the proposed scheme combining access control of attributes and blockchain technology in this paper can not only ensure the secure storage and integrity of medical information but also has a high throughput when accessing medical information.
\end{abstract}

\begin{IEEEkeywords}
	Medical Information, Hyperledger fabric, Smart contracts, ABAC, IPFS.
\end{IEEEkeywords}

\IEEEpeerreviewmaketitle

\section{Introduction}
\sethlcolor{red}
\IEEEPARstart{W}{ith}
the development of technology, various emerging technologies are merging with the healthcare sector, making the process of building healthcare information technology increasingly sophisticated \cite{sima2020influences}.
The World Health Organisation defines medical information as the most innovative and shareable asset.
Nowadays, the number of medical institutions around the world presents an index stage growth, and the medical data generated by medical institutions also present explosive growth. Due to the deepening of the degree of information in hospital information, the information system within the hospital gradually expands from a single HIS charging system into a system with electronic medical records. The medical data is accompanied by the registration, diagnosis, and hospitalization, medical data is gradually complex and stereochemical, and the importance of privacy and security is significantly increased \cite{schulz2019standards}.

\begin{figure}[htbp]
	\centering
	\includegraphics[width=\linewidth]{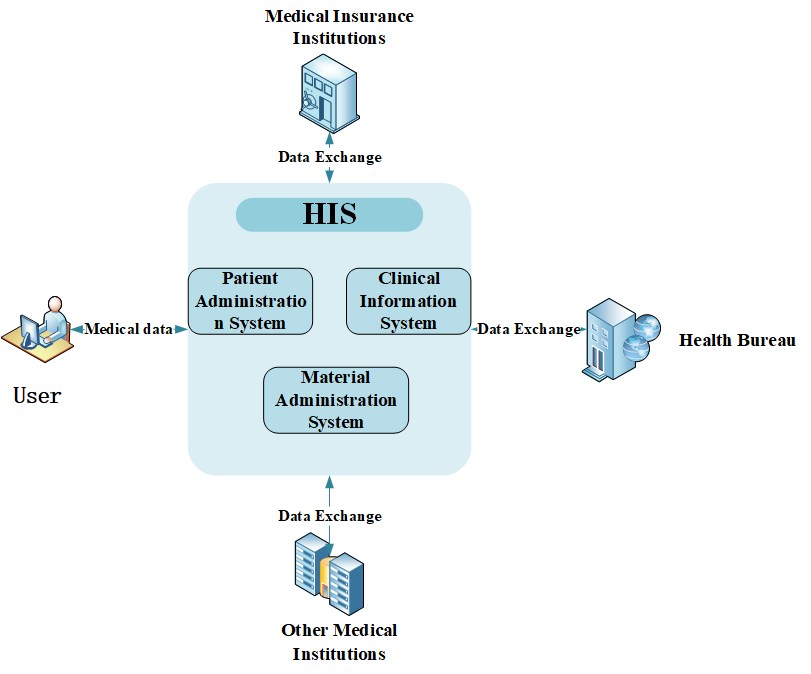}
	\caption{Medical data exchange}
	\label{fig: Medicaldata}
\end{figure}

Currently, the combination of traditional paper medical records and centralized medical data management systems is still the main form of medical institutions to store patients’ medical data as shown in Fig. \ref{fig: Medicaldata}.
However, this form of medical system faces severe risks of privacy disclosure \cite {xiang2021hadoop}.
Therefore, the transformation of the centralized medical data management system to distributed medical data sharing system is an irresistible trend of the whole society \cite{Li2021LDAB}.

However, since most medical and health institutions are isolated from each other, they store and maintain medical health data, forming data islands. This is not only not conducive to long-term records of patients with their disease development, but also caused a waste of medical equipment and a large number of medical health data resources duplication.
To maximize the value of medical health data, to meet the core needs of medical information construction, and provide more humanized and reasonable services for patients, sharing data between medical institutions is an inevitable trend \cite{xu2019healthchain}.
In addition, due to the extensive use of emerging Internet technology in the medical field, the medical data transmission methods and paths have become increasingly diversified, and gradually transferred from the internal transmission of hospitals to the transmission between medical institutions, medical institutions, and insurance and other institutions, and between patients and medical institutions, which also greatly increases the difficulty of patient data protection \cite{Zhang2019TransmissionLA}.
The above reasons lead to the characteristics of large scale, complex structure, and rapid growth of medical data, so it is difficult to find an ideal method to store medical information.

\sethlcolor{orange}
Fortunately, in recent years, the rise of blockchain technology has brought new solutions to the \hl{secure storage} of medical information.
In essence, blockchain is a distributed database with the characteristics of decentralization, security, and transparency \cite{Zheng2017AnOO, Fang2020rev, li2021blockchain}.
As a decentralized database, blockchain provides a reliable solution to the problems of poor sharing, low effectiveness, and weak security in medical data management. Data can be recorded on the real-time shared blockchain platform, and timestamps are added to ensure the immutability of the data. The tamper resistance of the blockchain ensures the security of medical data \cite{Li2021MFVTAnAT}. On the licensed blockchain, blockchain members can obtain data information through access operations.

Specifically, on the license blockchain, the blockchain member can obtain information of data by accessing operations, allowing the member to view outline information, to ensure sharing of medical data on a non-licked blockchain.
Mainstream blockchain projects can be divided into four categories:  cryptocurrency, platform, application, and asset token. Blockchain technology is widely used in smart cities \cite{Bhushan2020BlockchainFS, Sharma2018BlockchainBH},
Internet of things (IoT) \cite{Liu2020FabriciotAB, Wang2021IoT,Han2021ABA,yu2020IoT},
smart finance \cite{li2021fabric},
Internet of Vehicles (IoV) \cite{Wang2021IoV, Zhou2019BlockchainAC,Zhang2019ResearchOP,Cui2020BehaviorAA, Liu2020BlockchainBT},
and education \cite{AlHarthy2019TheUB, Oganda2020BlockchainES, Li2021MOOCsChainAB}.
Medical data involves personal privacy and sensitive information, such as personal name, ID number, and home address, so medical records become the primary goal of information theft, so it is urgent to combine blockchain technology and the medical sector.

Furthermore, blockchain has entered a new era with the emergence and continuous improvement of smart contracts and further development of blockchain projects such as Ether and Hyperledger. Smart contracts are programmable and Turing-complete \cite{Wang2019BlockchainEnabledSC}. Transactions can automatically initiate code based on rules set by the system, and the emergence of smart contracts has laid an important foundation for merging blockchain technology and medical information \cite{Li2020FabricChainC}.
In the open network environment of blockchain, the attribute-based access control(ABAC) model is a suitable and effective access control model. As a flexible fine-grained access control method \cite{Li2017FlexibleAF}, the model mainly determines that the data requester has the correct attributes to determine the data requester's access control authority to private data resources.
\sethlcolor{red}
So far, the application of blockchain technology in the medical field is not satisfactory. In this regard, we store medical data into blockchain by deploying intelligent contracts to ensure the privacy and security of medical data. At the same time, the ABAC model is introduced for access control to ensure that users can access them safely and efficiently. In addition, due to the huge and complex medical information, to alleviate the storage pressure of blockchain, we also combine the interstellar file system to realize the slimming of the whole blockchain and further improve the efficiency of user access.\hl{Compared with existing studies, the model proposed in this paper realizes more fine-grained access to medical information and at the same time alleviates the storage pressure of blockchain, making the throughput of the system greatly improved, which is also the advantage of this scheme.}

Specifically, the main contributions of this study are as follows.

\begin{itemize}
	\item{This paper applies blockchain to medical information management and realizes decentralized management and secure storage with the help of distributed consensus and authentication mechanisms.}
	\item {We design an auxiliary architecture based on ABAC, which can realize fine-grained access control and dynamic management of permissions.}
	\item {In this paper, we use smart contracts to define multi-tier data structures, access policies, and system workflows to improve the efficiency of data storage, retrieval, and query. }
	\item {We ease the storage pressure of blockchain with the interstellar file system. }
	\item {This paper designs simulation experiments and verifies the performance of the scheme.}
\end{itemize}

\sethlcolor{orange}
The rest of this article is as follows.
Section \ref{sec: Related Work} describes the related works.
In Section \ref{sec: Preliminaries}, we introduce the necessary background and technologies.
Next, \hl{Section} \ref{sec: System model and design} introduces the model, assumptions, and design objectives of the proposed scheme.
Then, Section \ref{sec: Experiment and comparison} sets up two groups of comparative experiments and then analyzes the results.
Finally, in \hl{Section} \ref{sec: Conclusion}, we summarize this paper and discuss further work.

\section{Related Work}
\label{sec: Related Work}
\sethlcolor{green}
In this section, we survey blockchain-based secure storage in section \ref{sec: Blockchain-based secure storage of medical data} and blockchain-based secure sharing in section \ref{sec: Blockchain-based secure sharing of medical data}.\hl{Although existing models and schemes achieve secure storage and sharing of medical information, they fail to realize fine-grained access to medical information, which will undoubtedly reduce the user experience. In addition, most existing studies have not considered the storage bottleneck of blockchain.}\sethlcolor{orange}
\hl{In order to make up for the deficiency of existing studies, this paper not only achieves the safe storage and sharing of medical information, but also optimizes the access control operation of medical information, and alleviates the storage pressure of blockchain to a certain extent, which is also the difference between the proposed scheme in this paper and the existing model.}

\subsection{Blockchain-based secure storage of medical data}
\label{sec: Blockchain-based secure storage of medical data}
The extension of blockchain technology to the healthcare field has a profound impact due to its decentralized, tamper-proof, and transparent nature.

Azaria \textit{et al.} \cite{Azaria2016MedRecUB} propose a decentralized blockchain-based MedRec system to handle EHR. MedRec has a modular design where the administrative privileges, authorization, and data sharing of the system are among the participants.
Medblock \cite{Fan2018MedBlockEA} is a hybrid architecture based on blockchain to protect EMR. The architecture nodes of the architecture are divided into endorsement nodes, sorting nodes, and submission nodes. The consensus algorithm used is a variant of the part consensus algorithm.
Conceição \textit{et al.} \cite{Conceio2018EletronicHR} propose a generic architecture for storing patient Electronic Health Record (EHR) data using blockchain technology.
%In \cite{Li2018BlockchainBasedDP}, a medical data storage system based on blockchain is proposed. The system can not only ensure the verifiability of medical data but also protect the privacy of patients.
Yang and Li \cite{Yang2018ADO} propose an EHR architecture based on blockchain. The architecture prevents tampering and abuse of EHR by tracking all events in the blockchain.
Kushch \textit{et al.} \cite{kushch2019blockchain} proposed a special data structure for storing electronic medical data on the blockchain:  blockchain tree. The structure of the blockchain tree is a sub-chain and one or more of a recorded patient identity and a sub-chain stored in additional critical information (such as diagnostic records), and blocks on the main chain are initial blocks of the sub-chain.

\subsection{Blockchain-based secure sharing of medical data}
\label{sec: Blockchain-based secure sharing of medical data}
In addition to safe storage, the blockchain is equally widely used in security sharing.
In medical record management, the application and research of the blockchain in the medical field have received much attention, and many research institutions around the world participate.

%Xue \textit{et al.} \cite{Zhu2019AnIM} propose a medical data sharing model based on the blockchain. This model solves network security issues by improving consensus mechanisms, allowing medical data to be inspected, saved, and synchronized between different medical institutions.
Xia \textit{et al.} \cite{Xia2017BBDSBD} proposed a blockchain-based system called men shared. The system can minimize the risk of data privacy and can be used to solve the problem of medical data sharing between medical data custodians in an untrusted environment.
Zhang \textit{et al.} \cite{Zhang2018TowardsSA} propose a blockchain-based medical data sharing scheme, which uses the private blockchain owned by the hospital to store the patient's health data, and uses the consortium blockchain to save the security index.
Zhang \textit{et al.} \cite{Zhang2018GenieAS} combined with artificial intelligence technology and blockchain technology proposed a safe and transparent medical data-sharing platform. This platform utilizes the transparency of the zone chain for data tracking, imparting the characteristics of non-tampered.
Liu \textit{et al.} \cite{Liu2018BPDSAB} use blockchain technology and cloud storage technology to propose a data-sharing scheme for paying attention to privacy protection in the medical field.
The scheme stores the original medical data in the cloud indexes the data in the blockchain and prevents the data from being maliciously modified by the tamper-proof feature of the blockchain.
%Chen \textit{et al.} \cite{chen2019blockchain} designed a storage scheme to manage personal medical blockchain-based and cloud-stored data that is not dependent on any third party and no party has absolute power to influence the processing.
%Ref. \cite{egala2021fortified} proposed a novel blockchain-based architecture that provides decentralized EHR and automation of services based on smart contracts without compromising system security and privacy.
To realize the dynamic communication between medical alliance chains,  Qiao \textit{et al.} \cite{Qiao2020DynamicAC} propose a scheme that allows dynamic communication between healthcare alliance chains, which enables patients to securely and autonomously share their records in an authorized healthcare alliance chain within milliseconds.

\section{Preliminaries}
\label{sec: Preliminaries}

This section mainly introduces the architecture of medical information security storage schemes based on blockchains and access controls.
Section \ref{sec: Blockchain technology in healthcare information storage} introduces the structure of the scheme,
section \ref{sec: Hyperledger Fabric} presents the workflow,
section \ref{sec: Attribute based access control model},
and section \ref{sec: IPFS} describes the smart contract design.

\subsection{Blockchain technology in healthcare information storage}
\label{sec: Blockchain technology in healthcare information storage}
Blockchain helps to build decentralized data sharing and application mechanisms. Traditionally, medical information management is a unilaterally maintained information system.
The drawback of this mode of management is that too centralized information management power makes it difficult to achieve real information sharing. Blockchain technology introduces the characteristics of distributed books.
Since the file information input under the blockchain technology is jointly maintained and supervised by multiple parties, the joint supervision of various information data by multiple departments ensures the openness and transparency of data information and also determines the openness and transparency of blockchain technology transactions rules \cite{Chen2018BlockchainBasedMR}. This will fundamentally solve the problems of low work efficiency and too chaotic a working state in traditional medical information management.

Moreover, blockchain can construct a credible deposit system. The management of medical archives information is nothing more than the four most basic processes of addition, deletion, modification, and query. However, in blockchain, the two basic processes of deletion and modification in archives information management are abandoned, and the process of archives information processing is reduced. The irreparability and security of data information in blockchain are guaranteed from the technical design. In addition, each block of information in the blockchain records the creation time and the hash value of the previous block. This chain structure marked with time itself facilitates the usual audit, tracking, and traceability, and improves the utilization rate of medical information.

Finally, blockchain can solidify data exchange and benefit allocation rules. The combination of intelligent contract and block link technology can maximize the automation of archival information sharing. Once the smart contract is implemented, it cannot stop and is not interfered with by external operations. Hospitals can use this feature to entrench interest distribution rules \cite{Egala2021FortifiedChainAB}. In medical information sharing, intelligent contracts can change the behavior of participants involved in information sharing into active participation, promote the efficiency and speed of information sharing, and truly maximize the value of medical information. In this compulsory information sharing, the secret box operation in traditional information sharing is constrained, and the quality of medical data information is ensured.

\subsection{Hyperledger Fabric}
\label{sec: Hyperledger Fabric}
\begin{figure}[htbp]
	\centering
	\includegraphics[width=\linewidth]{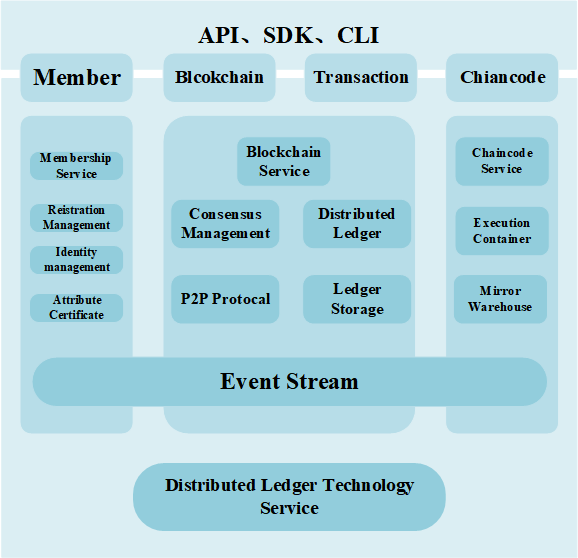}
	\caption{The module architecture diagram of Hyperledger fabric}
	\label{fig:fabric1}
\end{figure}

\begin{figure}[htbp]
	\centering
	\includegraphics[width=\linewidth]{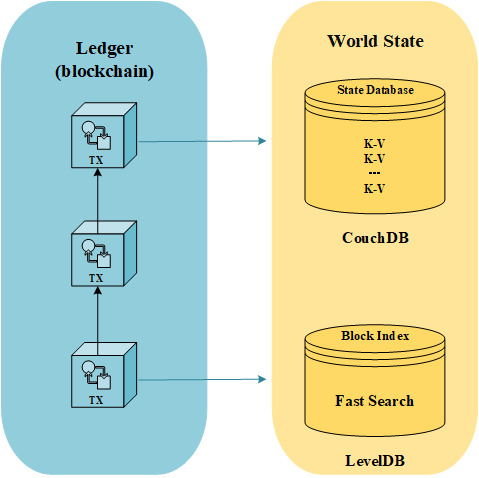}
	\caption{\hl{Structure of the ledger in Hyperledger Fabric.}}
	\label{fig:fabric2}
\end{figure}

In recent years, cryptocurrencies, represented by Bitcoin, have achieved great success, which has successfully drawn the world's attention to blockchain technology, however, such public chains have problems such as low transaction throughput, long transaction times, wasted resources, and data consistency. To address these issues the Linux Foundation created the Hyperledger project in 2015, which is one of the world's largest blockchain projects and is often used as a platform for enterprise blockchain development.
%allows authorized members to participate in data maintenance.
Hyperledger Fabric is designed with a modular architecture that includes members, blockchain, transactions, and smart contracts, as shown in Fig. \ref{fig:fabric1}.
\sethlcolor{red}
Member management module for the requirements of the enterprise-level blockchain to security and privacy, the member management module has strengthened the user's joining permissions, and anyone involved in the transaction needs to be certified by the PKI public key infrastructure. The blockchain module uses the P2P protocol to manage distributed books and can configure different consensus protocols according to different requirements, and record the transaction history in chain classification, with the latest state of the World State mechanism, the specific state of the ledger is specifically shown in Fig. \ref{fig:fabric2}.\hl{Hyperledger Fabric employs Apache Kafka(Distributed Messaging System) based on ZooKeeper(Distributed Services Framework). Kafka is essentially a message processing system where consumers of messages subscribe to specific topics and producers are responsible for publishing messages. In the whole Hyperledger Fabric network KafKa mainly provides transaction ordering service, that is, KafKa realizes the ordering service for all transaction requests in the network.}

\sethlcolor{green}

%\begin{figure}[htbp]
%	\centering
%	\subfigure[The module architecture diagram of Hyperledger fabric]{
%		\label{fig: fabric}
%		\includegraphics[width=0.45\linewidth]{pic/Fig_2.1}
%	}
%	\hspace{0.01\linewidth}
%	\subfigure[The module architecture diagram of Hyperledger fabric]{
%		\label{fig: accounts}
%		\includegraphics[width=0.45\linewidth]{pic/Fig_2.2}	
%	}
%	\caption{Diagram of Hyperledger fabric}
%	\label{fig: fabric12}
%\end{figure}

The transaction module controls the data in the transaction process in the form of deployment transactions and invocation transactions, where deployment transactions are installed on all peer nodes by Chaincode when the transaction is successfully executed, while invocation transactions are conducted by invoking the specified functions in the Chaincode through the SDK provided by the Fabric Software Development Kit.
Smart contracts record the business logic agreed by members of Fabric's federated chain and can be written in common languages such as Go and Java, overcoming the shortcomings of traditional blockchains that are limited to domain-specific languages.

\subsection{Attribute based access control model}
\label{sec: Attribute based access control model}
\begin{figure}[htbp]
	\centering
	\includegraphics[width=\linewidth]{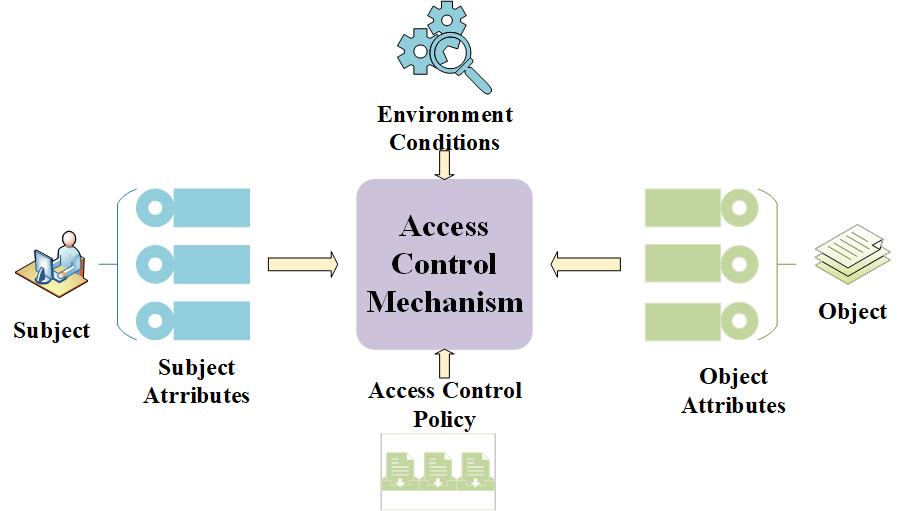}
	\caption{Structure of the ABAC model}
	\label{fig: ABAC}
\end{figure}

Attribute based access control is a comprehensive consideration of user, resource, operation, and contextual access control policies. It determines whether to grant access to the requester to configure the correct attribute, that is, this policy does not need to specify the relationship between the data requester and the private data, but by judging whether the data requester's attribute determines its pair access control permissions for this private data. Since the strategy is a more stable attribute due to the system operation.
Therefore, using the attribute to describe the access control policy to separate attribute management and access decision phase, and the specific implementation can increase or delete the policy according to the actual situation, implement the update modification of the policy, refine the access control particle size, and have good flexibility, sexuality and scalability.
Attributes are the core of the policy, which can be defined by a quadruplet $A \in \{S, O, P, E\}$, where each field has the following meaning:
$A$ represents attributes, each of which exists as a key-value pair.
$S$ represents subject attributes, including the subject's identity, role, position, and credentials.
$O$ represents object attributes, including the object's identity, location, department, type, data structure, etc.
$E$ represents the environmental attributes, including time, system status, security level, current access, etc.
$P$ represents the operation attributes, mainly used to describe the subject's access to the object type, such as write, modify, delete, etc. The structure of the model is shown in Fig. \ref{fig: ABAC}.
An attribute-based access control request (ABACR) can be defined as $\mathrm{ABACR}=\{\mathrm{AS} \wedge \mathrm{AO} \wedge \mathrm{AP} \wedge \mathrm{AE}\}$, where $AS$ represents the subject attribute, $AO$ represents the object attribute, $AP$ represents the operation attribute, and $AE$ represents the environment attribute.
$R$ represents a set of rules, which can also be defined by a quadruplet:  $\mathrm{R}\left(\mathrm{A}\left(\mathrm{S}_{\mathrm{i}}\right), \mathrm{A}\left(\mathrm{O}_{\mathrm{i}}\right), \mathrm{A}\left(\mathrm{E}_{\mathrm{i}}\right), \mathrm{A}\left(\mathrm{P}_{\mathrm{i}}\right)\right) \rightarrow$ \{Allow,Deny$\}$, this formula indicates that the subject authorizing attribute $S_i$ at the time of access performs an access action $P_i$ on object $O_i$ in a contextual environment with attribute value $E_i$, with two outcomes i.e. allow or deny.

\subsection{InterPlanetary File System (IPFS)}
\label{sec: IPFS}
Designed by Juan Benet and developed by Protocol Labs with the help of the open-source community since 2014, IPFS (Interstellar File System) is a network transfer protocol designed to create persistent and distributed storage and sharing of files. IPFS combines features of existing technologies, including DHT, BitTorrent, Git, and SFS, to achieve the primary function of storing data locally and connecting nodes to each other for data transfer.
IPFS was originally designed to build a better resource network than the now commonly used HTTP protocol to compensate for the shortcomings of HTTP. Compared to HTTP, IPFS exhibits advantages such as fast download speeds, global storage, security, and data perpetuation. IPFS is essentially a content-addressable, versioned, peer-to-peer hypermedia distributed storage and transport protocol. It has the following features.
Content Addressable: IPFS only cares about the content of the file, generating a unique hash mark from the file content, which is accessed by the unique mark and checked in advance to see if the mark has already been stored. If it has been stored, it is read directly from other nodes, without the need for duplicate storage, saving space in a sense.
Slicing large files: files placed in IPFS nodes do not care about their storage path or name. IPFS provides the ability to slice and dice large files, downloading multiple slices in parallel when used.
Decentralized, distributed network structure:  Such a network structure is suitable for solving bottlenecks in the blockchain's storage capacity by storing large amounts of hypermedia data on IPFS.
Encrypted storage: IPFS adds a cryptographic hash unique to digital information to the encrypted data, and the corresponding hash of the stored file cannot be changed. The hash corresponds to the file one-to-one.
In an IPFS network, there is no need to take into account the location of the server and the name and path of the file. When a file is placed in an IPFS node, each file is given a unique hash value calculated based on its contents. When access to a file is requested, IPFS finds the node where the file is located based on the hash table and fetches the file. IPFS combined with blockchain can be a good solution to the blockchain storage problem.

\section{Experimental Methods}
\label{sec: System model and design}
This section mainly introduces the architecture of medical information security storage schemes based on blockchains and access controls.
Section \ref{sec: System architecture} introduces the structure of the scheme, section \ref{sec: Workflow} presents the workflow, and section \ref{sec: Smart contract of the scheme} describes the smart contract design of the scheme.

\subsection{Scheme architecture}
\label{sec: Scheme architecture}
\begin{figure}[htbp]
	\centering
	\includegraphics[width=\linewidth]{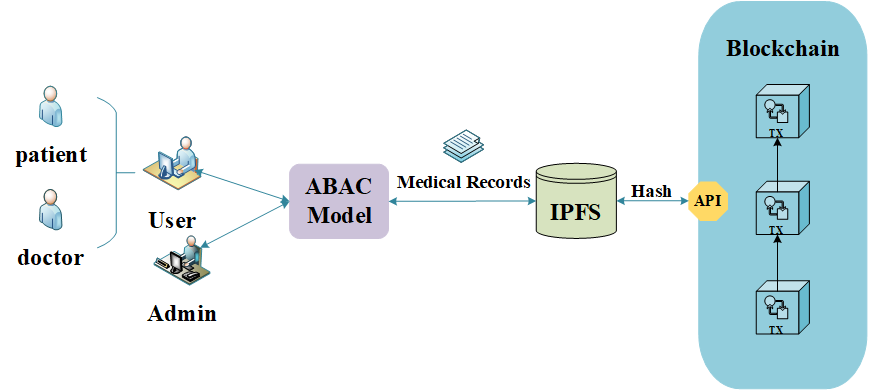}
	\caption{\hl{The proposed scheme architecture}}
	\label{fig: system}
\end{figure}

The architecture of the system consists of a user, an attribute-based access control model, an interstellar file system, and a blockchain, the detailed architecture of which is shown in Fig. \ref{fig: system}.
\sethlcolor{green}

Users can be divided into two types: normal users, who can be doctors and patients, both of whom can participate in the authorization of the solution and thus access medical data. Administrator users are responsible for managing the blockchain and can create or update smart contracts.
The combination of the ABAC model and this scheme forms a medical information access control model. The specific description of the model is as follows.

\begin{equation}
	\label{eq: 1}
	\begin{aligned}
		P=\{AS, AO, AP,AE\}
	\end{aligned}
\end{equation}

\begin{equation}
	\label{eq: 2}
	\begin{aligned}
		AS=\{ userId, role, department \}
	\end{aligned}
\end{equation}

\begin{equation}
	\label{eq: 3}
	\begin{aligned}
		AO=\{recordId\}
	\end{aligned}
\end{equation}

\begin{equation}
	\label{eq: 4}
	\begin{aligned}
		AP= \begin{cases}1, & allow  \\ 0, & deny \end{cases}	\end{aligned}
\end{equation}

\begin{equation}
	\label{eq: 5}
	\begin{aligned}
		AE=\{ createTime, endTime \}
	\end{aligned}
\end{equation}

Policy(P): It represents the access control policy based on attributes contains four elements in the set, namely AS, AO, AP, and AE.

Attribute of Subject(AS): It includes three main types of attributes, namely user ID (identifies the unique identity of the user), user role (doctor and patient), and user department (specific department).

Attribute of Object(AO): It includes the medical record ID (identifies the uniqueness of the record).

Attribute of Permission(AP): An attribute that indicates whether the user has access to the medical record, with 1 representing permission and 0 representing denial.

Attribute of Environment(AE): The environmental conditions required for the access control policy, mainly including the creation time (when the policy was created) and the end time (when the policy expires). If the current time of a policy is later than the end time, it means that the policy is invalid.
\sethlcolor{red}
IPFS: It is mainly used to mitigate the storage pressure of the blockchain. The medical data stored in IPFS will be stored in a \hl{MerkleDAG} to ensure the security of the data, which is called the address hash. Then, the address hash is stored into the zone chain, thereby replacing the original data.
In IPFS, the original data is subjected to the SHA256 algorithm twice and then Base58 encoding, resulting in a hash length of 33 Bytes.
So the original medical information is replaced with the hash address, which will greatly reduce a block.
The size of the whole blockchain is achieved.

Blockchain: The blockchain is the heart of the solution, a distributed network of trusted nodes that ensures the synchronization and storage of medical data, thus ensuring data security and accuracy. In this solution the blockchain is developed based on Hyperledger Fabric and access control can be implemented by writing smart contracts.

\subsection{Workflow}
\label{sec: Workflow}
\begin{figure}[htbp]
	\centering
	\includegraphics[width=\linewidth]{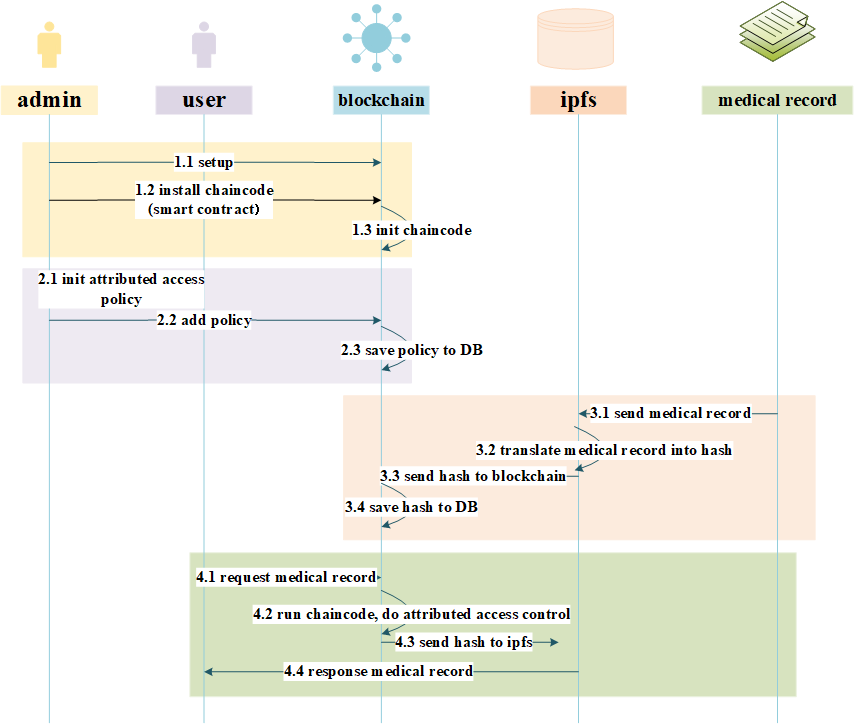}
	\caption{The workflow of the proposed scheme}
	\label{fig: workflow}
\end{figure}

The workflow of the proposed scheme mainly contains four parts. This section describes each part, and the specific workflow is shown in Fig. \ref{fig: workflow}.
The symbols used are shown in Tab. \ref{tab: acronyms}.

\begin{table}[htbp]
	\centering
	\caption{The summary of acronyms \& definitions}
	\label{tab: acronyms}
	\resizebox{\linewidth}{!}{
		\begin{tabular}{l|l}
			\hline
			\hline
			\textbf{Notations} & \textbf{Discription}                                    \\
			\hline
			CA                 & Certificate Authority                                   \\
			Cert               & Certificate file                                        \\
			conf               & Config file of the node                                 \\
			F(x)...            & Functions defined in source code                        \\
			CC                 & Chaincode in Hyperledger Fabric                         \\
			ASC                & Access Smart Contract                                   \\
			PSC                & Policy Smart Contract                                   \\
			RSC                & Record Smart Contract                                   \\
			Image              & Docker Image                                            \\
			Container          & Docker Container                                        \\
			TX                 & Transaction in blockchain                               \\
			AS,AO,AP,AE        & Attributes of subject,object,permission,and environment \\
			Ledger             & Ledger in Hyperledger Fabric                            \\
			SDB                & State Database in Hyperledger Fabric                    \\
			IPFS               & InterPlanetary File System                              \\
			Cli                & Blockchain system client                                \\
			ABACP              & Attribute based Access Control Policy                  \\
			\hline
			\bottomrule
	\end{tabular}}
\end{table}

\subsubsection{Part 1}
The basic procedure of this program is the installation of the construction and Chaincode of the blockchain network. These basic processes need to be completed by the administrator user. Process 1 is mainly divided into three steps as follows.

\textbf{Step 1}:  Prior to building a specific blockchain network, all members of the network must register the certificate and the required certificate is issued by CA.

\begin{equation}
	\label{eq: 6}
	\begin{aligned}
		CA \rightarrow \left\{Cert_{peer}, Cert_{orderer}, Cert_{ channel}, Cert_{user} \right\}
	\end{aligned}
\end{equation}

All peer nodes and orderer nodes run in docker containers and the relevant certificates they require need to be packaged into a docker image before they can be run.

\begin{equation}
	\label{eq: 7}
	\begin{aligned}
		Build(conf, Cert) \stackrel{buid}{\longrightarrow} Image \stackrel{run}{\longrightarrow} Container
	\end{aligned}
\end{equation}

After setting up all the peer and orderer nodes, start creating channels, each in a separate blockchain and ledger as

\begin{equation}
	\label{eq: 8}
	\begin{aligned}
		\{blockchain, ledger\} \stackrel{join}{\longrightarrow} Channel
	\end{aligned}
\end{equation}

\textbf{Step 2}:  After the above operation, a basic blockchain network has been built, and the Chaincode is written next to create an application.

\begin{equation}
	\label{eq: 9}
	\begin{aligned}
		Code(F{x} \ldots) \rightarrow CC
	\end{aligned}
\end{equation}

The administrator user uses the Hyperledger Fabric SDK or Client to install the Chaincode, and all peer nodes must have the Chaincode installed.

\begin{equation}
	\label{eq: 10}
	\begin{aligned}
		Install(CC)\stackrel{SDK/Client}{\longrightarrow} Peer	\end{aligned}
\end{equation}

\textbf{Step 3}:  Once the chaincode is completed, we need to initialize it by calling the invoke function to complete the initialization of the chaincode, and the initialized Chaincode is stored in the container.

\begin{equation}
	\label{eq: 11}
	\begin{aligned}
		Invoke(Init) \stackrel{S D K / Client} {\longrightarrow Peer}
	\end{aligned}
\end{equation}

\subsubsection{Part 2}
This section requires the specification of relevant access control policies and the whole process needs to be agreed upon between the user and the administrator. The policy needs to be saved to the blockchain by the administrator once it has been created.

\textbf{Step 1}:  Administrators and users set access control policies based on $AS$, $AO$, $AP$, and $AE$.

\begin{equation}
	\label{eq: 12}
	\begin{aligned}
		\operatorname{Decide}(A S, A O, A P, A E) \rightarrow A B A C P
	\end{aligned}
\end{equation}

\textbf{Step 2}:  The administrator uploads the developed access control policy to the blockchain network.

\begin{equation}
	\label{eq: 13}
	\begin{aligned}
		\mathrm{Upload}(\mathrm{ABACP}) \rightarrow \text { Contract }
	\end{aligned}
\end{equation}

\textbf{Step 3}:  The administrator runs PSC to implement operations such as adding and modifying policies and saves the final policy values to the SDB and ledger.

\begin{equation}
	\label{eq: 14}
	\begin{aligned}
		\mathrm{PSC}(\mathrm{ABACP}) \rightarrow\{\mathrm{SDB}, \text { ledger }\}
	\end{aligned}
\end{equation}

\subsubsection{Part 3}
This section implements the storage of medical information by first uploading the medical records into IPFS to get a hash address, and then saving that address to the blockchain.

\textbf{Step 1}:  Users upload medical records to IPFS.

\begin{equation}
	\label{eq: 15}
	\begin{aligned}
		\text { Upload(Medical Record) } \rightarrow \text { IPFS }
	\end{aligned}
\end{equation}

\textbf{Step 2}:  IPFS translates medical records into a hash address according to its operational mechanism.

\begin{equation}
	\label{eq: 16}
	\begin{aligned}
		\text { IPFS(Medical Record) } \stackrel{\text { translate }}{\longrightarrow} \text { hash }
	\end{aligned}
\end{equation}

\textbf{Step 3}:  Send the hash address to the blockchain.

\begin{equation}
	\label{eq: 17}
	\begin{aligned}
		\text { Send(hash) } \rightarrow \text { blockchain }
	\end{aligned}
\end{equation}

\textbf{Step 4}:  Save medical information to the SDB and ledger by running the smart contract RSC.

\begin{equation}
	\label{eq: 18}
	\begin{aligned}
		\operatorname{Run}(\mathrm{RSC}) \rightarrow\{\mathrm{SDB}, \text { ledger }\}
	\end{aligned}
\end{equation}

\subsubsection{Part 4}
This section is a process for accessing medical information based on attribute access control and can be divided into four specific steps.

\textbf{Step 1}:  The user initiates a request for access to medical data.

\begin{equation}
	\label{eq: 19}
	\begin{aligned}
		\text { Request } \rightarrow \text { blcokchain }
	\end{aligned}
\end{equation}

\textbf{Step 2}:  Upon receipt of a user request, the ASC contract is called to verify that the user has access to the data.

\begin{equation}
	\label{eq: 20}
	\begin{aligned}
		\operatorname{ASC}(\text { Request }) \rightarrow \begin{cases}1, & \text { allow } \\ 0, & \text { deny }\end{cases}
	\end{aligned}
\end{equation}

\textbf{Step 3}:  If the user has access rights, then the blockchain transfers the hash of the medical information to the IPFS.

\begin{equation}
	\label{eq: 21}
	\begin{aligned}
		\text { Blockchain(hash) } \stackrel{\text { send }}{\longrightarrow} \text { IPFS }
	\end{aligned}
\end{equation}

\textbf{Step 4}:  IPFS calculates the medical data requested by the user based on the hash address.

\begin{equation}
	\label{eq: 22}
	\begin{aligned}
		\text { Response(Medical Record) } \rightarrow \text { Cli }
	\end{aligned}
\end{equation}

\subsection{Smart contract of the scheme}
\label{sec: Smart contract of the scheme}
Smart contracts are not only related to the implementation of access control, but also the storage of medical information, and are therefore at the heart of this solution. There are three smart contracts in total:  policy contract (PSC), access control contract (ASC), and medical record contract(RSC).

\subsubsection{Policy Contract (PSC)}
The PSC provides the following methods to manipulate ABACPs.

$CheckPolicy()$: PSC needs to verify the validity of the ABACP by this method. Each ABACP should contain $AS$, $AO$, $AP$, and $AE$, and all the four attributes should be satisfied for this policy to be valid.
%The details are shown in Algorithm \ref{al:PSC.CheckPolicy()}.

$AddPolicy()$: The PSC needs to run the $CheckPolicy()$ method before calling this method to add the policy, and only after the policy is legal can the policy be written to SDB and blockchain. The details are shown in Algorithm \ref{al:PSC.AddPolicy()}.

$DeletePolicy()$: This method will be called in two ways. Firstly, the administrator will call this method to delete an ABACP. Secondly, when the $CheckAccess()$ method is executed and a policy is found to have expired, then this method will be called automatically to delete the useless policy. This is shown in Algorithm \ref{al:PSC.DeletePolicy()}.

$UpdatePolicy()$: This method is called when an administrator needs to modify an ABACP. This method is called when the administrator needs to modify an ABACP. The modification record is also written to the SDB and the blockchain. This method also executes the $AddPolicy()$ method at the end after the policy is updated, adding the modified policy back to the blockchain.

$QueryPolicy()$:  all policies are stored in the state database CouchDB (a kind of key-value pair database) and the administrator can query the details of the desired ABACP by using the property $AS$ or $AO$.

%\begin{algorithm}[htbp]
%	\caption{PSC.CheckPolicy()}
%	\label{al:PSC.CheckPolicy()}
%	\begin{algorithmic}[1]
%		\Require
%		ABACP
%		\Ensure
%		True or False
%		\State  \{AS, AO, AP, AE\} ← ABACP
%		\State State = True
%		\For {item in AS}
%		\If {item $\in$ \{userId, role, department\}}
%		\State State = False
%		\EndIf
%		\EndFor
%		\For {item in AO do}
%		\If {item $\in$ \{recordId, userId\}}
%		\State State = False
%		\EndIf
%		\EndFor
%		\If {Val(AP) != 1 or 0}
%		\State State = False
%		\EndIf
%		\For {item in AE}
%		\If {item $\in$ \{createTime, endTime\}}
%		\State State = False
%		\EndIf
%		\EndFor
%		\State \Return State
%	\end{algorithmic}
%\end{algorithm}

\begin{algorithm}[htbp]
	\caption{PSC.AddPolicy()}
	\label{al:PSC.AddPolicy()}
	\begin{algorithmic}[1]
		\Require
		ABACP
		\Ensure
		Ok or Error
		\State APIstub ChaincodeStub ← Invoke()
		\If{CheckPolicy(ABACP) == False}
		\sethlcolor{red}
		\State \Return \hl{Error(BadPolicy)}
		\EndIf
		\State {AS, AO} ← ABACP
		\State ABACP$_{id}$ ← HASH$_{sha256}$(AS + AO)
		\State err ← A APIstub.PutState(ABACP$_{id}$, ABACP)
		\If{err! = null}
		\sethlcolor{red}
		\State \Return \hl{Error}
		\EndIf
		\State \Return Ok
	\end{algorithmic}
\end{algorithm}

\begin{algorithm}[htbp]
	\caption{PSC.DeletePolicy()}
	\label{al:PSC.DeletePolicy()}
	\begin{algorithmic}[1]
		\Require
		AS, AO
		\Ensure
		Ok or Error
		\State APIstubChaincodeStub ← Invoke()
		\State PolicyID ← HASH$_{sha256}$(AS + AO)
		\State err ← APIstub.GetState(Id)
		\If{err! = null}
		\State \Return Error
		\EndIf
		\State err ← APIstub.DelState(PolicyID)
		\If{err! = null then}
		\State \Return Error
		\EndIf
		\State \Return Ok
	\end{algorithmic}
\end{algorithm}

\subsubsection{Access Control Contract (ASC)}
ASC primarily implements the access control function, i.e. determining whether a user's access control-based request matches the prescribed access control policy.
The methods in ASC are as follows.
$CheckAccess()$: This method is the core of the implementation of access control, as shown in Algorithm \ref{al:ASC.CheckAccess()}.
If the method returns a null result, it proves that there is no policy that supports the request and the request is invalid.
If the result is not null, it means that there is a policy that matches the request. Finally, the request is verified by validating the eligible policy, and if the attributes $AE$ and $AP$ in the policy are both satisfied, the request is proven to pass the verification.

\begin{algorithm}[htbp]
	\caption{ASC.CheckAccess()}
	\label{al:ASC.CheckAccess()}
	\begin{algorithmic}[1]
		\Require
		ABAC\_Request
		\Ensure
		Ok or Error
		\State {A$_{u}$S, A$_{u}$O, A$_{u}$E} ← GetAttrs(ABAC\_Request)
		\State  P ← PSC.QueryPolicy(A$_{u}$S, A$_{u}$O)
		\If{P == Null}
		\State \Return Error();
		\EndIf
	    \State	\{ . . . , A$_{p}$P, A$_{p}$E\} ← P
	
		\If{Value(A$_{p}$P) == 1 \&\& A$_{p}$E.endTime \textgreater  currentTime}
		\State \Return Ok;
		\EndIf
		\State \Return Error()
	\end{algorithmic}
\end{algorithm}

\subsubsection{Policy Contract (PSC)}
The RSC is primarily used to store a hash address representing a complete medical record.
The user first uploads the medical record to IPFS, which then returns a hash address for the medical record, which is then uploaded to the blockchain and SDB.
$AddRecord()$: This method stores the hash address from IPFS into the blockchain, i.e. the key-value pair $<recordId, hash>$ into the SDB. The details are shown in Algorithm.\ref{al:RSC.AddRecord()}.

\begin{algorithm}[htbp]
	\caption{RSC.AddRecord()}
	\label{al:RSC.AddRecord()}
	\begin{algorithmic}[1]
		\Require
		Medical Record(MR)
		\Ensure
		Ok or Error
		\State APIstubChaincodeStub ← Invoke()
		\State IPFS$_{hash}$ ← IPFS(MR)
		\State RecordID ← HASH$_{sha256}$(MR.recordId)
		\State err ← APIstub.PutState(RecordID, IPFS$_{hash}$)
		\If{err! = null}
		\State \Return Error
		\EndIf
		\State \Return Ok
	\end{algorithmic}
\end{algorithm}

$DeleteRecord()$: This method first deletes the hash address from the SDB, and then deletes the complete medical record from the IPFS based on $recordId$.
$UpdateRecord()$: When this method is executed, it first updates the medical data in the IPFS to get a new hash address, and then restores this new hash address to the SDB by calling the $AddRecord()$ method.
$QueryRecord()$: This method first looks up the hash address of the medical record in the SDB based on the $recordId$, and then sends the found hash address to IPFS to be parsed into a complete medical record.

\section{Experiment and Results}
\label{sec: Experiment and comparison}
This section introduces the process of the experiment and the final results, which are used to verify the performance of this solution through comparison.
Section \ref{sec: Experimental environment} introduces the experimental environment, i.e. the hardware and software resources required for the experiment.
Section \ref{sec: Creation and realization} introduces the process of creating and implementing the solution.
Section \ref{sec: Experimental results and comparisons} presents the experimental results, which are used to compare and analyze the performance of the solution.

\subsection{Experimental environment}
\label{sec: Experimental environment}
The hardware and software resources required for the stand-alone environment for this solution are shown in Tab. \ref{tab: Resources}.

\begin{table}[htbp]
	\centering
	\caption{Hardware and software resources}
	\label{tab: Resources}
	\begin{tabular}{l|l}
		\toprule
		\hline
		\multicolumn{2}{l}{\textbf{Hardware}}   \\
		\hline
		CPU                & i7 7500u 2.9GHz    \\
		Memory             & 8G                 \\
		Hard Disk          & 1T                 \\
		\hline
		\multicolumn{2}{l}{\textbf{SoftWare}}   \\
		\hline
		OS                 & Ubuntu Linux 16.04 \\
		Docker             & v19.03.2           \\
		Docker-compose     & v1.24.1            \\
		Node               & v12.12.0           \\
		Golang             & v1.15.8            \\
		Hyperledger Fabric & v1.4.6             \\
		\hline	
		\bottomrule
	\end{tabular}
\end{table}

\subsection{Creation and realization}
\label{sec: Creation and realization}
This section mainly includes three parts,
section \ref{sec: Network architecture and initialization process} mainly introduces the network structure of the scheme and initialization configuration and start.
Section \ref{sec: Chaincode installation and upgrade} introduces the installation of the Chaincode.
Section \ref{sec: System implementation} mainly introduces how to use the attribute-based access control model to call intelligent contracts.

\subsubsection{Network architecture and initialization process}
\label{sec: Network architecture and initialization process}
The scheme consists of a total of eight network nodes,
%as shown in Tab. \ref{tab: Num},
and the steps for network initialization are shown below.
%
%\begin{table}[htbp]
%	\centering
%	\caption{Number of network nodes}
%	\label{tab: Num}
%	\resizebox{\linewidth}{!}{
%		\begin{tabular}{l|l|l}
%			\hline
%			\hline
%			\textbf{Node name} & \textbf{Discription}       & \textbf{Number} \\
%			\hline
%			Fabric-couchdb     & Database node              & 4               \\
%			Fabric-ca          & CA node                    & 2               \\
%			Fabric-peer        & peer node                  & 4               \\
%			Fabric-orderer     & orderer node               & 1               \\
%			Fabric-tools       & cli node                   & 1               \\
%			PSC                & policy smart contract node & 4               \\
%			ASC                & access smart contract node & 4               \\
%			RSC                & record smart contract node & 4   \\
%			\hline
%			\bottomrule
%	\end{tabular}}
%\end{table}

\textbf{Step 1}: Use cryptogen tools to generate organization structure and identity certificates for your network.

\textbf{Step 2}: Use the configtxgen tool to generate the creation block for Orderer, the configuration transaction file for the channel, and the anchor node configuration update file for each organization.

\textbf{Step 3}: First start the fabrics network with docker-compose, then use client nodes to create channels, and finally add each peer node to the channels.

\subsubsection{Chaincode installation and upgrade}
\label{sec: Chaincode installation and upgrade}
Firstly, installation. After the initialization of the blockchain network, the chaincode can be installed. The chaincode is installed through the hyperledger client node. The client node is used to install the chaincode into each peer node in turn.
Secondly, instantiation. After installing the chaincode, specify any peer node to instantiate the installed chain code.
Finally, upgrade. Before updating the chain code, you must install the new chain code, that is, the chain code update is only valid on the peer node with the new chain code installed.

\subsubsection{System implementation}
\label{sec: System implementation}
In Hyperledger Fabric users can access the blockchain via a client or an SDK, in this scenario a client written by the SDK will be used to interact with the blockchain.
The specific steps are as follows.

\textbf{Step 1}: The CA node generates a key pair for the client, which is stored in the user's wallet.

\textbf{Step 2}: The administrator connects the client to the peer node, and once the link is complete, the transaction can be submitted or evaluated.

\begin{figure}[htbp]
	\centering
	\includegraphics[width=\linewidth]{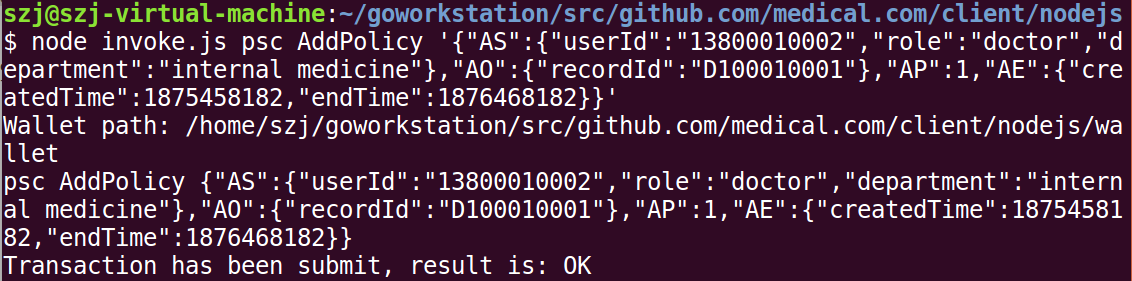}
	\caption{The result of calling the $PSC.AddPolicy()$ method}
	\label{fig: PSCadd}
\end{figure}

\textbf{Step 3}: First the orderer node completes the sorting process, then a consensus is reached between the peer nodes, and finally the status database can be queried or updated.
If you want to add a policy, you can call the $AddPolicy()$ method in PSC, as shown in Fig. \ref{fig: PSCadd}.

\begin{figure}[htbp]
	\centering
	\includegraphics[width=\linewidth]{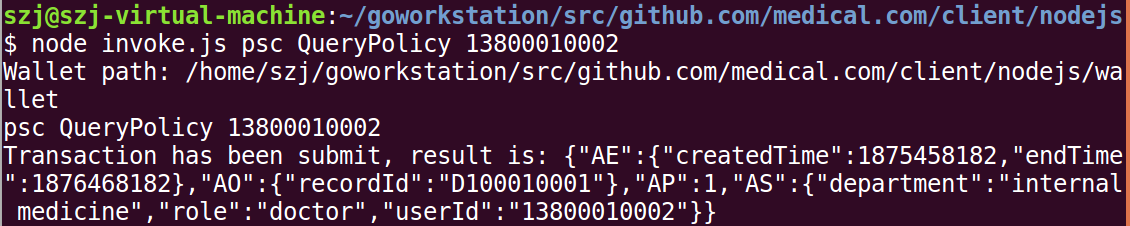}
	\caption{Results of calling the $PSC.QueryPolicy()$ method}
	\label{fig: PSCqu}
\end{figure}

As shown in Fig. \ref{fig: PSCqu}, if you want to know whether a policy has been added successfully, you can call the $QueryPolicy()$ method in the PSC to query the details of a policy.

\begin{figure}[htbp]
	\centering
	\includegraphics[width=\linewidth]{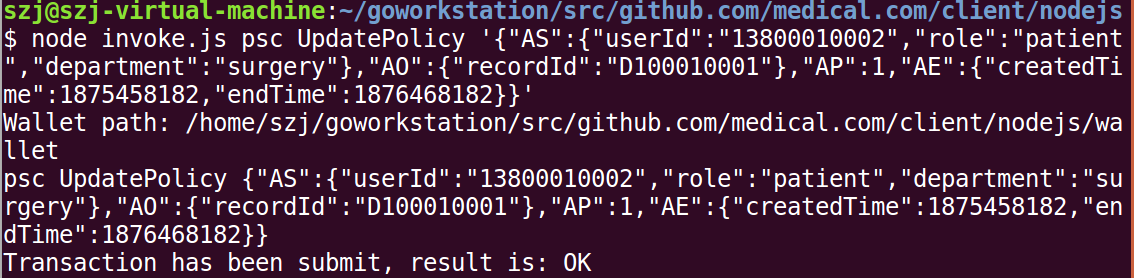}
	\caption{Results of calling the $PSC.UpdatePolicy()$ method}
	\label{fig: PSCUpate}
\end{figure}

As shown in Fig. \ref{fig: PSCUpate}, this policy can be updated by calling the $UpdatePolicy()$ method in the PSC for some reason to adapt to the new case.

\begin{figure}[htbp]
	\centering
	\includegraphics[width=\linewidth]{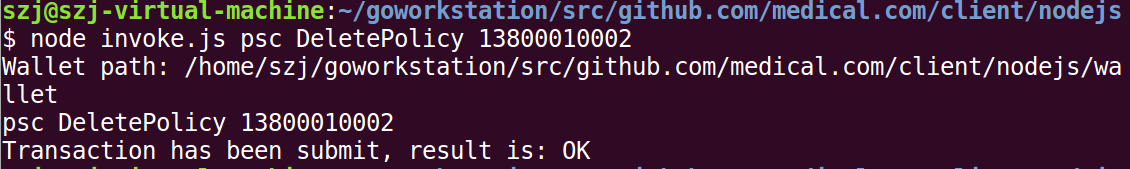}
	\caption{Results of calling the $PSC.DeletePolicy()$ method}
	\label{fig: PSCDelete}
\end{figure}

If a policy becomes invalid or the administrator needs to force the deletion of a policy, the policy can be deleted by calling the $DeletePolicy()$ method in the PSC. This is shown in Fig. \ref{fig: PSCDelete}.

\begin{figure}[htbp]
	\centering
	\includegraphics[width=\linewidth]{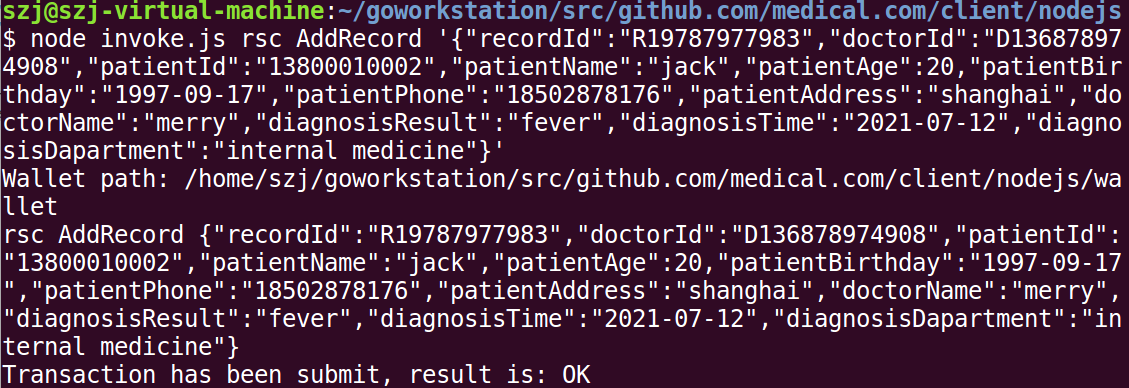}
	\caption{Results of calling the $RSC.AddRecord()$ method}
	\label{fig: RSCAdd}
\end{figure}

\begin{figure}[htbp]
	\centering
	\includegraphics[width=\linewidth]{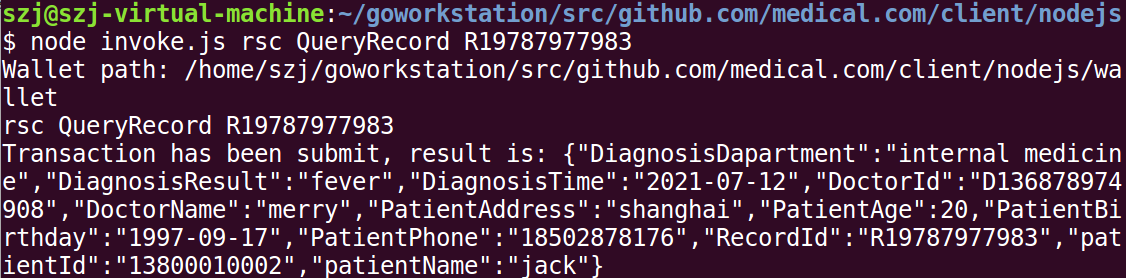}
	\caption{The result of calling the $RSC.QueryRecord()$ method}
	\label{fig: RSCQu}
\end{figure}

As shown in Fig. \ref{fig: RSCAdd}, if the Medical Centre needs to add a new medical record, it can do so by calling the $AddRecord()$ method in the RSC.
As shown in Fig. \ref{fig: RSCQu}, if the medical center needs to query the details of a medical record, it can do so by calling the $QueryRecord()$ method in the RSC.
If a medical record needs to be adjusted in real-time due to a new change in the patient's condition, the $UpdateRecord()$ method in the RSC can be called to update a medical record. If a medical record needs to be deleted due to age or other reasons, it can be deleted by calling the $DeleteRecord()$ method in the RSC.

\begin{figure}[htbp]
	\centering
	\includegraphics[width=\linewidth]{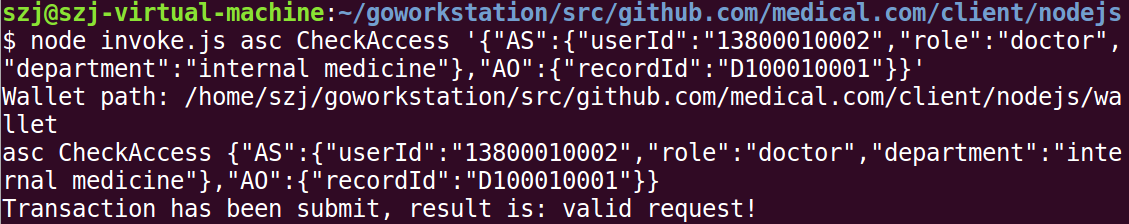}
	\caption{Results of calling $ASC.CheckAccess()$ method}
	\label{fig: ASCcheck}
\end{figure}

After receiving the user's request, it will automatically call the $CheckAccess()$ method in ASC to verify whether the request is reasonable. If the request is reasonable, it will return the ' valid request ! ', otherwise the request is invalid. The details are shown in Fig. \ref{fig: ASCcheck}.

\subsection{Results and discussion}
\label{sec: Experimental results and comparisons}
\begin{figure*}[htbp]
	\centering
	\subfigure[Time spent in PSC with different concurrent requests]{
		%	\label{fig: queryUser}
		\includegraphics[width=0.80\linewidth]{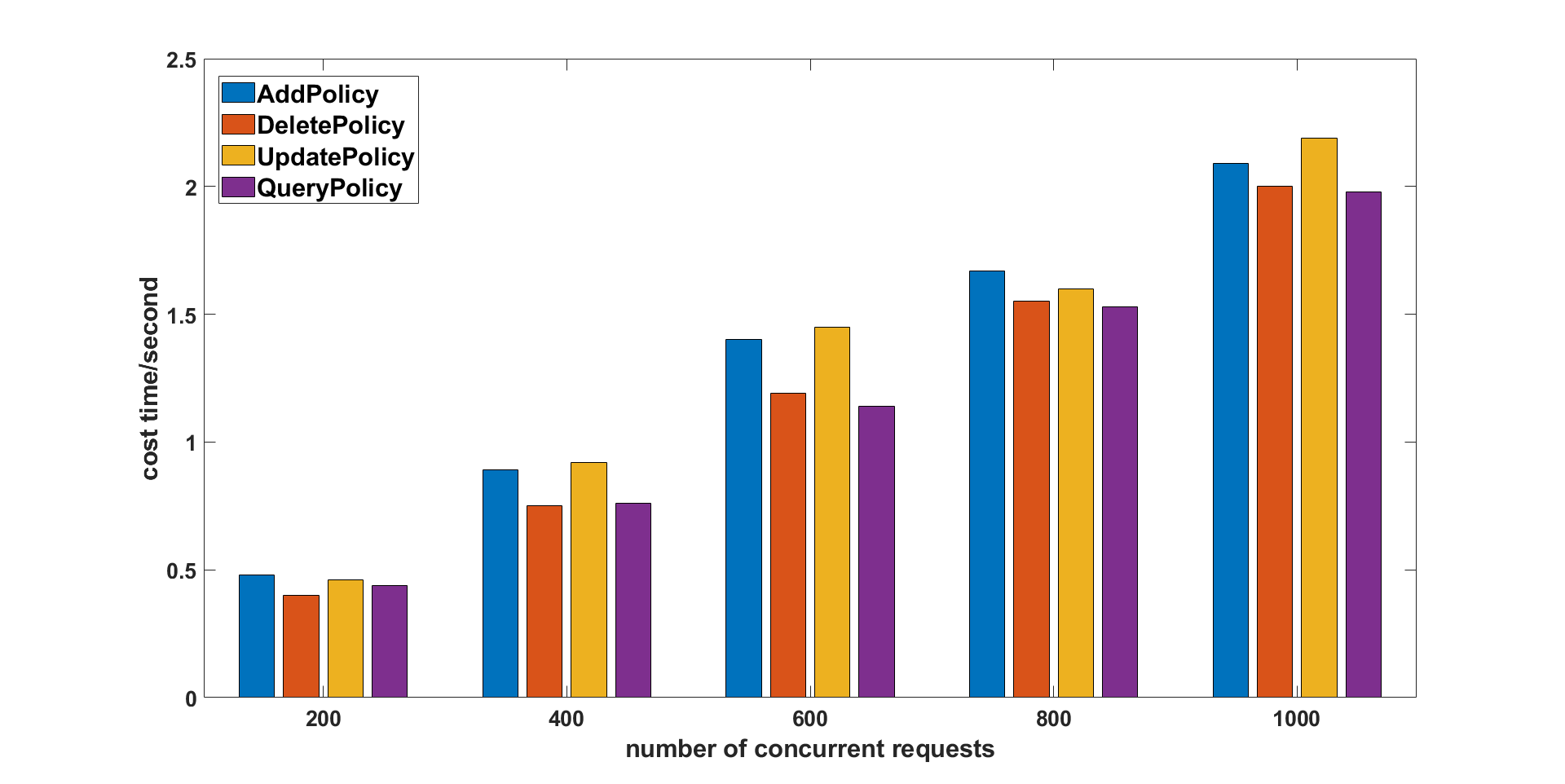}
	}
	\hspace{0.01\linewidth}
	\subfigure[Time spent by RSC under different concurrent requests]{
		%	\label{fig: queryProject}
		\includegraphics[width=0.80\linewidth]{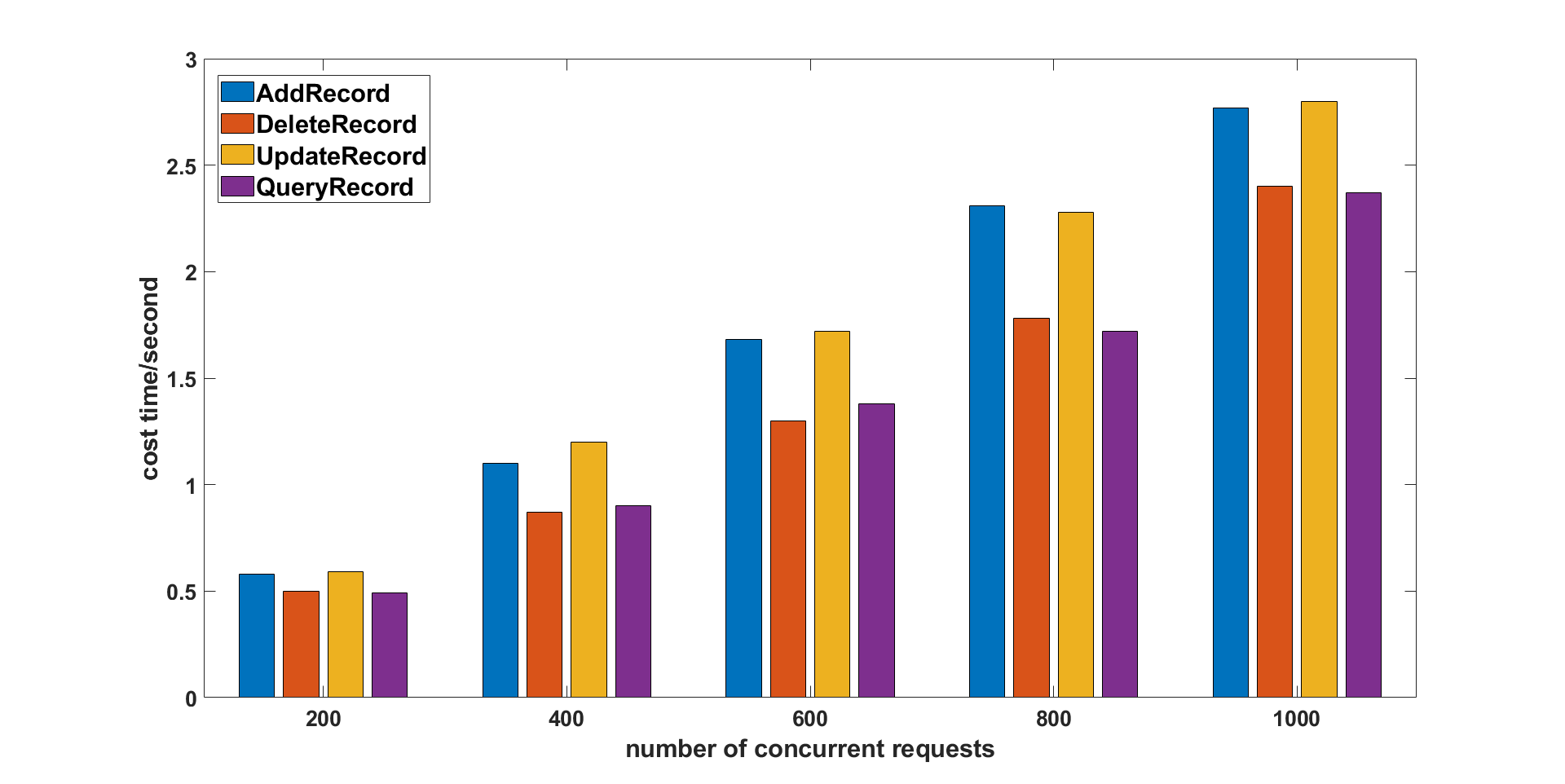}
	}
	%\vfill
	\subfigure[Time spent by ASC under different concurrent requests]{
		%	\label{fig: queryTask}
		\includegraphics[width=0.80\linewidth]{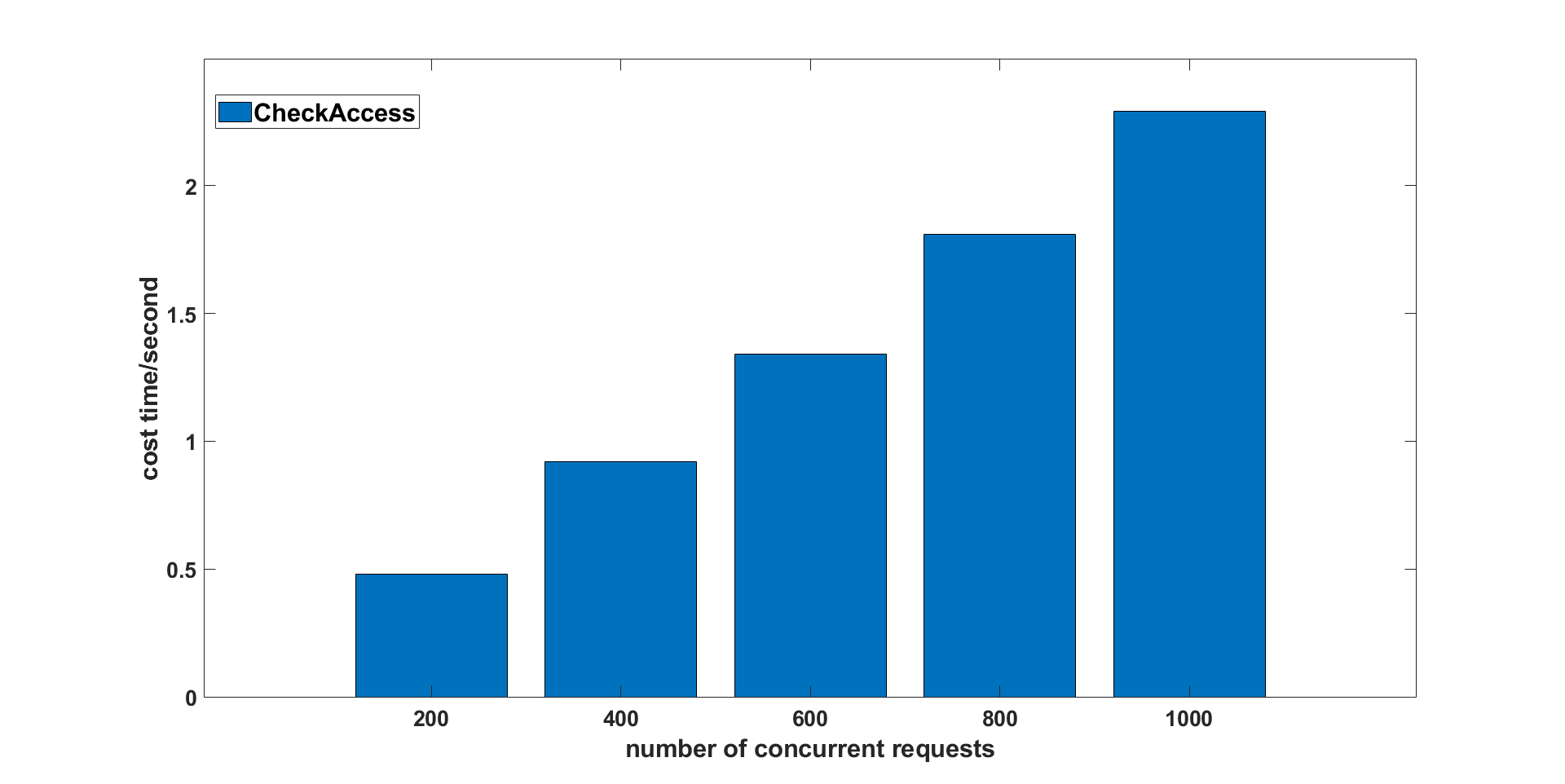}
	}
	%\hspace{0.01\linewidth}
	\caption{\hl{Time spent on different concurrent requests}}
	\label{fig: performance}
\end{figure*}

\begin{figure*}[htbp]
	\centering
	\subfigure[Throughput of PSC under different concurrent requests]{
		\label{fig: TPSC}
		\includegraphics[width=0.80\linewidth]{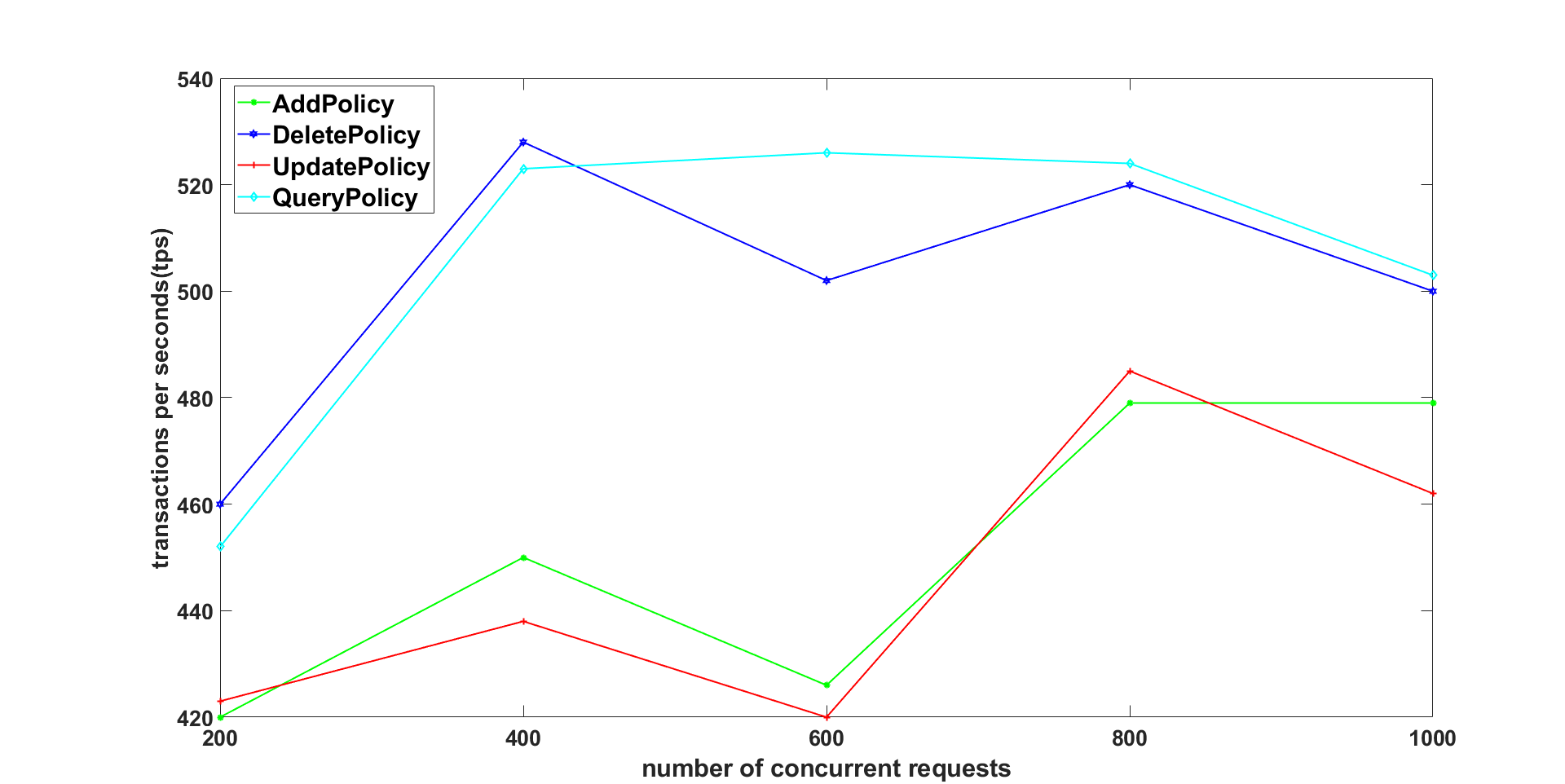}
	}
	\hspace{0.01\linewidth}
	\subfigure[Throughput of RSC under different concurrent requests]{
		\label{fig: TRSC}
		\includegraphics[width=0.80\linewidth]{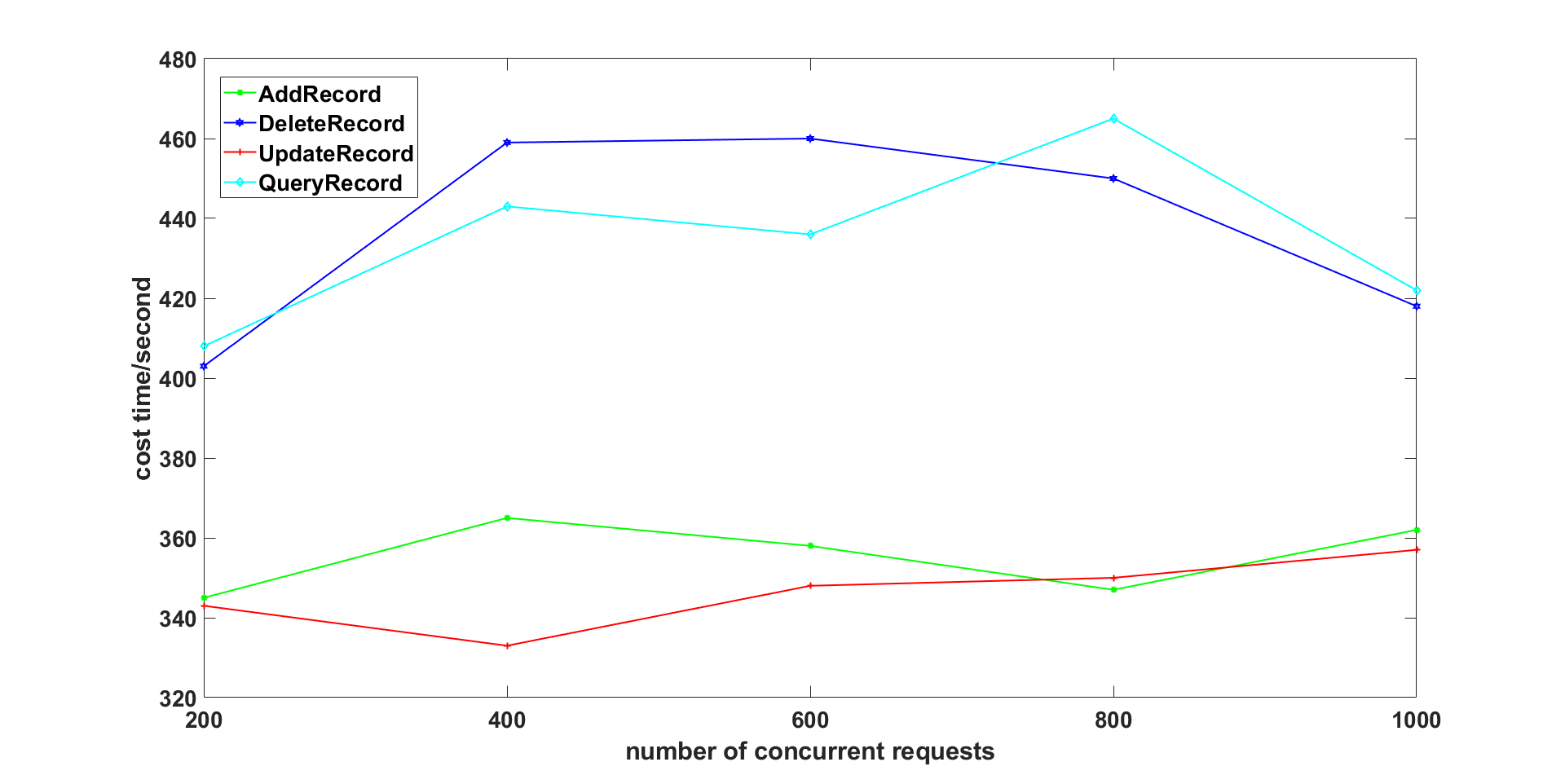}	
	}
   % \hspace{0.01\linewidth}
    \subfigure[Throughput of ASC under different concurrent requests]{
	%	\label{fig: changeTask}
	    \includegraphics[width=0.80\linewidth]{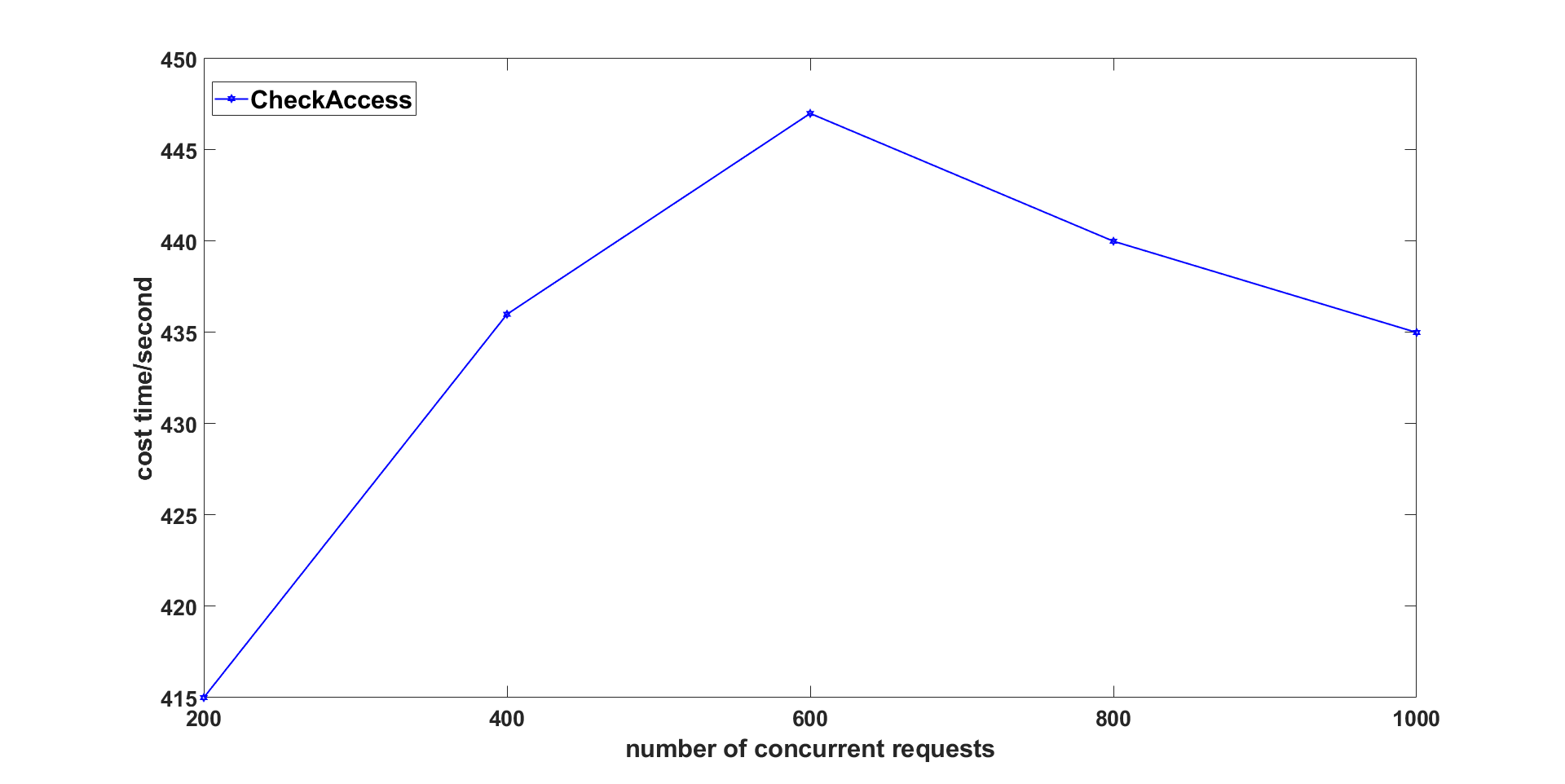}}
	\caption{\hl{Throughput under different concurrent requests}}
	\label{fig: TPS}
\end{figure*}

To verify the performance of this scheme, two sets of experiments are designed. In the first set of experiments, the number of virtual concurrent clients is set to 200, 400, 600, 800, and 1000, showing the time and throughput of PSC, ASC and RSC in dealing with transactions under different concurrent requests.
The details are shown in Fig. \ref{fig: performance} and Fig. \ref{fig: TPS}.

\sethlcolor{green}

\begin{figure}[htbp]
	\centering
	\includegraphics[width=0.85\linewidth]{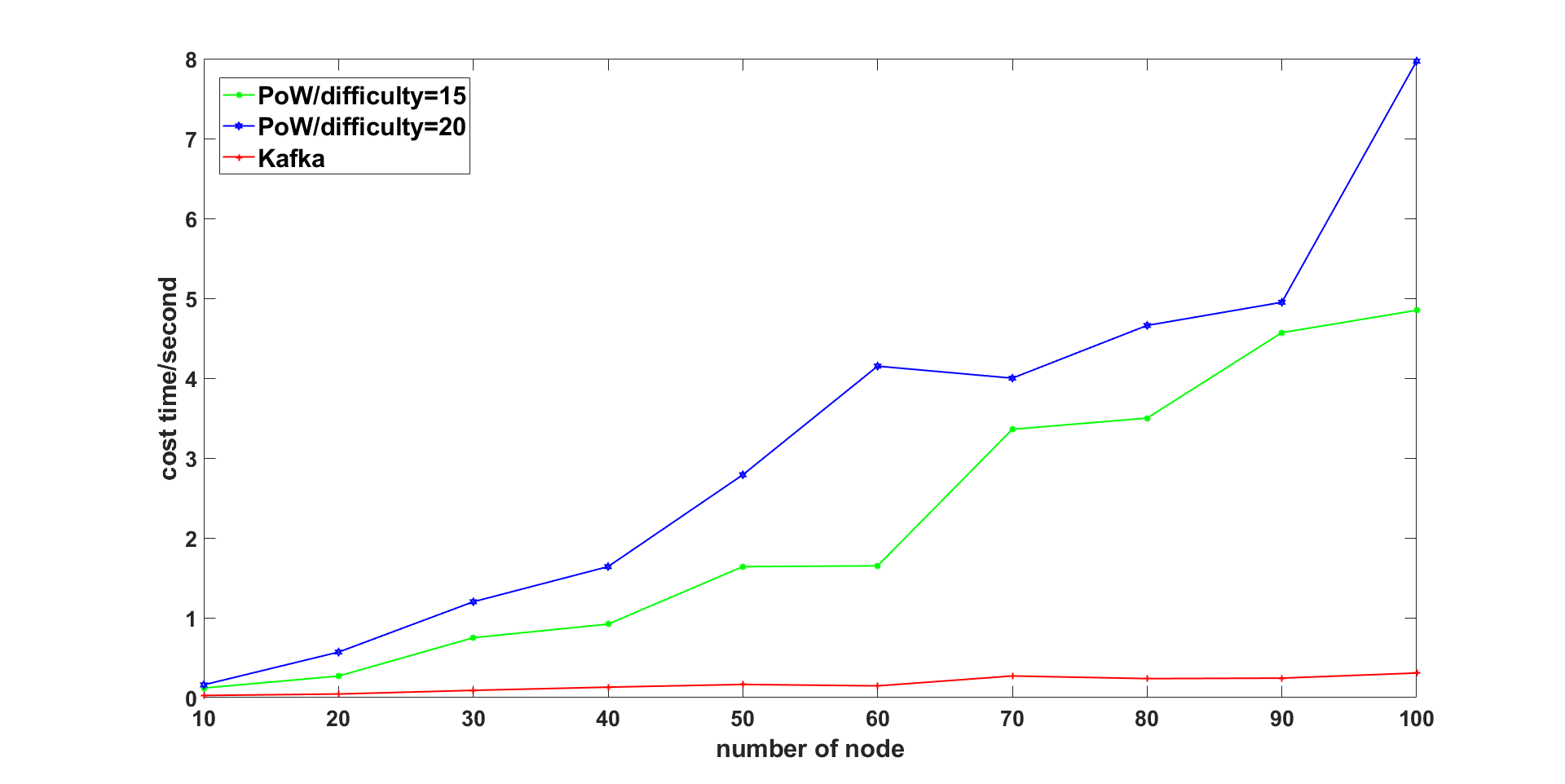}
	\caption{\hl{Comparison of Kafka and PoW in the time of reaching consensus}}
	\label{fig: consensus}
\end{figure}

\sethlcolor{orange}
The following conclusions are drawn from the above experimental results:
Firstly, add and update operations take a longer time, while query and delete operations take less time.
Secondly, the throughput of add and update operations is less than that of query and delete operations.
The throughput does not decrease significantly when the number of concurrent requests reaches a certain value.
\hl{Although PoW consensus can achieve complete decentralization, taking too long to reach consensus results in a large waste of resources. However, Kafka consensus can not only accomplish high throughput of transactions, but also provide sufficient fault tolerant workspace for consensus and ordering services.}
As shown in Fig. \ref{fig: consensus}, in the second group of experiments, we compared the differences in consensus time between the Kafka consensus mechanism and the PoW consensus mechanism adopted in this scheme by setting the number of different nodes (between 10 and 100).
The results show that this scheme can reach a consensus in a short time.
Through the above two groups of experiments, it can be proved that this scheme can not only maintain high throughput in a large-scale request environment but also effectively reach consensus in a distributed system.
\sethlcolor{green}

\section{Conclusion}
\label{sec: Conclusion}
This paper combines blockchain technology with an attribute-based access control model to take full advantage of blockchain technology to break down information silos in medical data and safeguard the security and privacy of medical information. In addition, the interstellar file system is utilized in storage to ease the storage pressure on the blockchain. The scheme uses a distributed architecture to achieve dynamic fine-grained access. The deployment and invocation of the chain code are described in detail and proof is given through experiments. In conclusion, this paper provides a practical reference for related research and can provide ideas for researchers.
Future work could make improvements in the following areas.

\begin{enumerate}
	\item This scheme is carried out on a single PC, future consideration could be given to using clusters to further optimize the performance of the distributed system.
	\item This scheme is based on the consensus mechanism of Kafka. To further reduce the arithmetic power and improve the consensus efficiency, a combination of other consensus algorithms can be considered in the future, such as using a consensus approach that combines Byzantine fault-tolerant algorithms with non-Byzantine fault-tolerant algorithms.
	\item This paper combines IPFS and blockchain to alleviate the storage pressure of blockchain, but this is only a transitional stage, and in the future, we should consider solving the data storage problem from the blockchain.
\end{enumerate}

\appendices

\section*{Declaration of interests}
\subsection*{Availability of data and materials}
Data sharing is not applicable to this article as no datasets were generated or analyzed during the current study.

\subsection*{Competing Interest}
The authors declare that they have no known competing financial interests or personal relationships that could have appeared to influence the work reported in this paper.

\subsection*{Funding}
This research is supported by the National Natural Science Foundation of China under Grant 61873160, Grant 61672338 and Natural Science Foundation of Shanghai under Grant 21ZR1426500.

\subsection*{Authors Contribution}
ZS proposed and developed the new idea of the paper and drafted it. HD and LD have substantially revised it. WX and WZ conducted the data analysis and text combing. CC is responsible for supervision.
All authors approved the submitted version. All authors read and approved the final manuscript.

\section*{List of Abbreviations}
\begin{table}[htbp]
	\centering
	\caption{List of Abbreviations}
	\label{tab: Abbreviations}
	\resizebox{\linewidth}{!}{
		\begin{tabular}{l|l}
			\hline
			\hline
\textbf{Abbreviations} & \textbf{Description}                   \\
\hline
ABAC                   & Attribute-Based Access Control         \\
IPFS                   & InterPlanetary File System             \\
HIS                    & Hospital Information System            \\
EHR                    & Electronic Health Record               \\
EMR                    & Electronic Medical Record              \\
PKI                    & Public Key Infrastructure              \\
P2P                    & Peer-to-Peer                           \\
SDK                    & Software Development Kit               \\
ABACR                  & Attribute-Based Access Control Request \\
DHT                    & Distributed Hash Table                 \\
SFS                    & Scalable File Service                  \\
HTTP                   & Hyper Text Transfer Protocol           \\
AS                     & Attributes of subject                  \\
AO                     & Attributes of object                   \\
AP                     & Attributes of permission               \\
AE                     & Attributes of environment              \\
MerkleDAG              & Merkle directed acyclic graph          \\
SHA256                 & Secure Hash Algorithm 256              \\
CA                     & Certificate Authority                  \\
Cert                   & Certificate file                       \\
Conf                   & Config file of the node                \\
F(x)...                & Functions defined in source code       \\
CC                     & Chaincode in Hyperledger Fabric        \\
ASC                    & Access Smart Contract                  \\
PSC                    & Policy Smart Contract                  \\
RSC                    & Record Smart Contract                  \\
Image                  & Docker Image                           \\
TX                     & Transaction in blockchain              \\
SDB                    & State Database in Hyperledger Fabric   \\
Cli                    & Blockchain system client               \\
ABACP                  & Attribute based Access Control Policy                 \\
			\hline
			\bottomrule
	\end{tabular}}
\end{table}
\ifCLASSOPTIONcaptionsoff
\newpage
\fi



% http: //mirror.ctan.org/biblio/bibtex/contrib/doc/
% http: //www.michaelshell.org/tex/ieeetran/bibtex/
%\bibliographystyle{IEEEtran}

\bibliographystyle{IEEEtran}
\bibliography{Ref}

\section*{Figure Title and Legend}
\begin{itemize}
	\item Figure 1:Medical data exchange
	\item Figure 2:The module architecture diagram of World state
	\item Figure 3:The module architecture diagram of World state
	\item Figure 4:Structure of the ABAC model
	\item Figure 5:The proposed system architecture
	\item Figure 6:The workflow of the proposed scheme
	\item Figure 7:The result of calling the $PSC.AddPolicy()$ method
	\item Figure 8:Results of calling the $PSC.QueryPolicy()$ method
	\item Figure 9:Results of calling the $PSC.UpdatePolicy()$ method
	\item Figure 10:Results of calling the $PSC.DeletePolicy()$ method
	\item Figure 11:Results of calling the $RSC.AddRecord()$ method
	\item Figure 12:The result of calling the $RSC.QueryRecord()$ method
	\item Figure 13:Results of calling $ASC.CheckAccess()$ method
	\item Figure 14:Time spent on different concurrent requests   (a)Time spent in PSC with different concurrent requests   (b) Time spent by ASC under different concurrent requests  (c) Throughput under different concurrent requests   (d) Throughput  of ASC under different concurrent requests.
	\item Figure 15: Throughput under different concurrent requests  (a) Throughput of PSC under different concurrent requests  (b) Throughput under different concurrent requests
	\item Figure 16: Comparison of Kafka and pow in the time of reaching consensus
\end{itemize}

\begin{IEEEbiography}[{\includegraphics[width=1in,height=1.25in,clip,keepaspectratio]{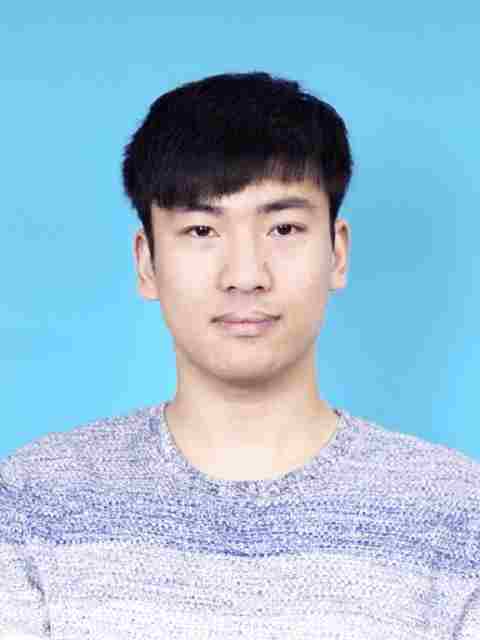}}]{Zhijie Sun} received the B.S degree in information management and information system from Henan Polytechnic University. and he is currently pursuing the M.S. degree in Shanghai Maritime University. His main research interests include blockchain technology and its applications and cryptography.
\end{IEEEbiography}

\begin{IEEEbiography} [{\includegraphics[width=1in,height=1.25in,clip,keepaspectratio]{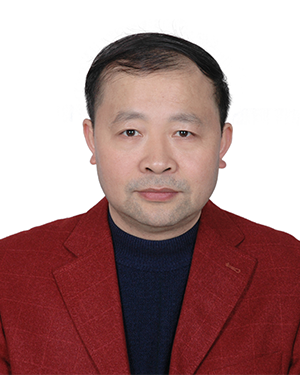}}] {Dezhi Han}
	received the BS degree from Hefei University of Technology, Hefei, China, the MS  degree and PhD degree from Huazhong University of Science and Technology, Wuhan, China. He is currently a professor of computer science and engineering at Shanghai Maritime University. His specific interests include storage architecture, Blockchain technology, cloud computing security and cloud storage security technology.
\end{IEEEbiography}

\begin{IEEEbiography}[{\includegraphics[width=1in,height=1.25in,clip,keepaspectratio]{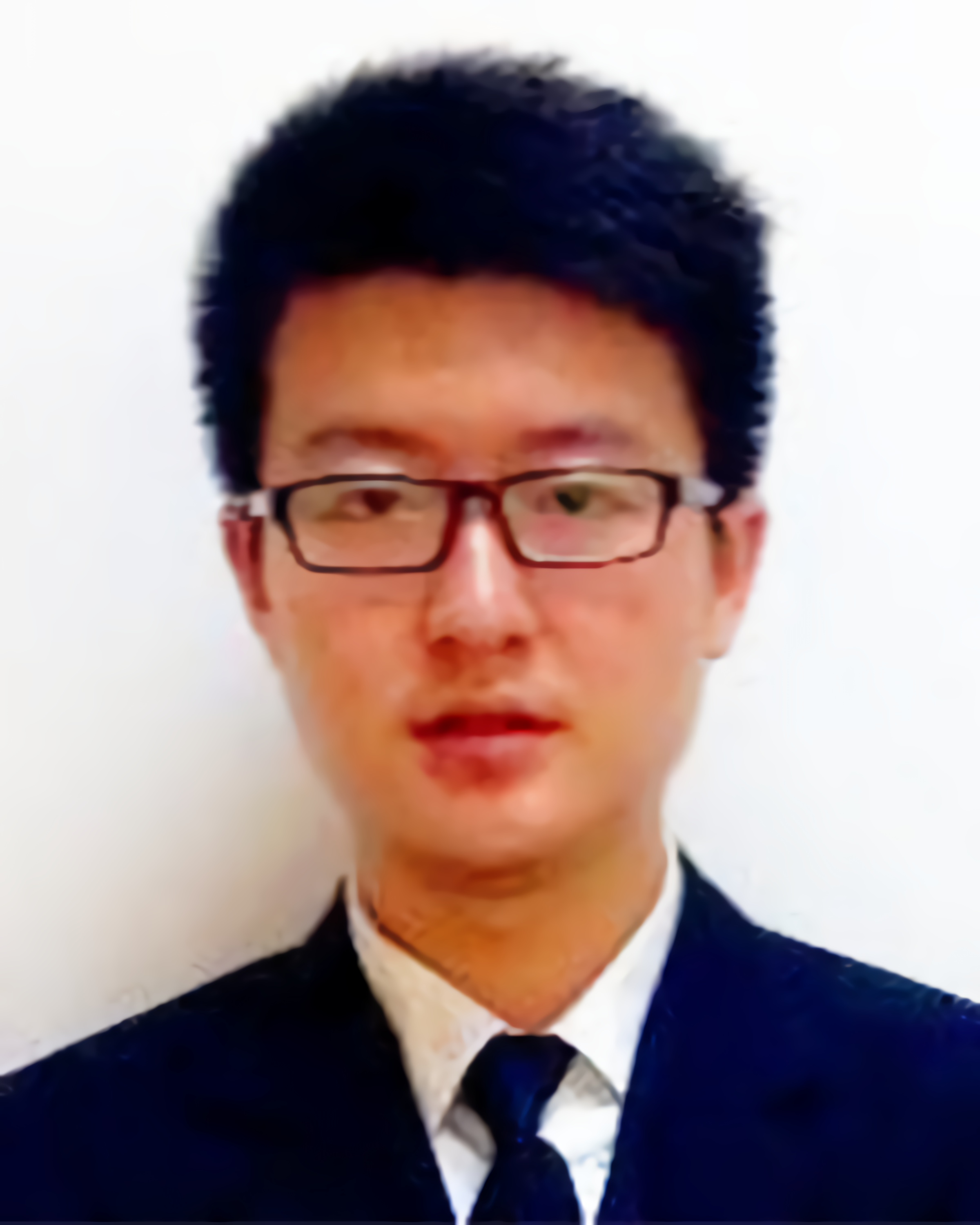}}]{Dun Li} received the B.S. degree in Human Resource Management from the Huaqiao University, Quanzhou, China, in 2013,  and the M.S. degree in Finance from the Macau University of Science and Technology, Macau, China, in 2015. He is currently doing his Ph.D. degree in Information Management and Information Systems at Shanghai Maritime University. His main research interests include smart finance, big data, machine learning, IoT, and Blockchain.
\end{IEEEbiography}

\begin{IEEEbiography}[{\includegraphics[width=1in,height=1.25in,clip,keepaspectratio]{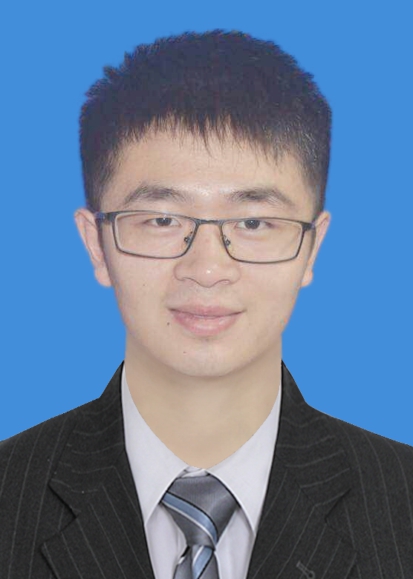}}]{Xiangsheng Wang} is currently pursuing the PhD at Shanghai Maritime University. His main research interests include natural language processing, image processing, visual question answering and machine learning.	
\end{IEEEbiography}

\begin{IEEEbiography}[{\includegraphics[width=1in,height=1.25in,clip,keepaspectratio]{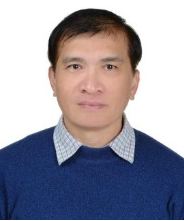}}]{Chin-Chen Chang} received the Ph.D. degree in computer engineering from National Chiao Tung University, Hsinchu, Taiwan, in 1982, and the B.E. and M.E. degrees in applied mathematics, computer and decision sciences from National Tsinghua University, Hsinchu, Taiwan, in 1977 and 1979, respectively. He was with National Chung Cheng University, Minxiong, Taiwan . Currently, he is a Chair Professor with the Department of Information Engineering and Computer Science, Feng Chia University, Taichung, Taiwan, since 2005. His current research interests include database design, computer cryptography, image compression, and data structures. Prof. Chang was a recipient of many research awards and honorary positions by and in prestigious organizations both nationally and internationally, such as the Outstanding Talent in Information Sciences of Taiwan. He is currently a Fellow of the IEEE, a Fellow of the IEE, U.K and a Member of the IEICE.
\end{IEEEbiography}

\begin{IEEEbiography}[{\includegraphics[width=1in,height=1.25in,clip,keepaspectratio]{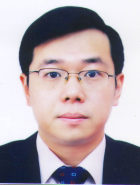}}]{Zhongdai Wu}, male, born in August 1976, is the Deputy General Manager, Chief Engineer, and Chief Information Officer of COSCO Shipping Technology Company Limited, with a doctoral degree, a researcher-level senior engineer, and a senior information manager of SASAC. He has more than 20 years of experience in shipping and logistics informatization construction. He has been responsible for the construction of various large-scale informatization projects of central enterprises and has presided over the completion of one Shanghai Key New Product, one Shanghai High-tech Achievement Transformation Project, one Shanghai Application Demonstration Project, and many software copyrights, and has rich experience in project planning and management. He has published more than ten academic papers, two of which were indexed by EI. He has been awarded as one of the top ten civilizational pacesetters of China Shipping Group, Shanghai New Long March Pioneer, Shanghai Federation of Trade Unions Scientific and Technological Innovation Talent, State-owned Assets Supervision, and Administration Commission Central Enterprise Knowledge-based Advanced Worker, Shanghai Young Post Leader, etc. Main research areas: shipping informationization, container management, ship and cargo management, logistics and supply chain technology research, cloud data center construction and management, network security situational awareness, shipping e-commerce, Internet of Things application, business intelligence technology, shipping big data application, ship satellite communication, etc.
\end{IEEEbiography}

%\newpage


%\vfill


\end{document}